\documentclass[aps,pra,pdflatex,reprint,10pt,showpacs,superscriptaddress,floatfix]{revtex4-2}
\usepackage{float}
\usepackage{amsmath}
\usepackage[T1]{fontenc}
\usepackage[utf8]{inputenc}
\usepackage{lmodern}
\usepackage{amssymb}
\usepackage{graphicx}
\usepackage{hyperref}
\usepackage{epstopdf}
\usepackage{pifont}
\usepackage{float}
\usepackage{textgreek}
\usepackage{ulem}
\usepackage{placeins}
\usepackage{tikz}
\usepackage{physics}
\usepackage{orcidlink}

\usepackage{multirow}

\graphicspath{{pdf/}}
\hypersetup{colorlinks=true,linkcolor=blue,citecolor=blue,filecolor=blue,urlcolor=blue,breaklinks=true}
\RequirePackage{color}

\usepackage{float}
\begin{document}

\date{\today}

\title{Torsion-induced confinement and tunable nonlinear optical gain in a mesoscopic electron system}

\author{Carlos Magno O. Pereira\orcidlink{0000-0002-5170-7538}}
\email[Carlos Magno O. Pereira - ]{carlos.mop@discente.ufma.br}
\affiliation{Programa de P\'{o}s-Gradua\c{c}\~{a}o em F\'{i}sica, Universidade Federal do Maranh\~{a}o, 65080-805, S\~{a}o Lu\'{i}s, Maranh\~{a}o, Brazil}

\author{Edilberto~O~Silva\orcidlink{0000-0002-0297-5747}}
\email{edilberto.silva@ufma.br}
\affiliation{Programa de P\'{o}s-Gradua\c{c}\~{a}o em F\'{i}sica, Universidade Federal do Maranh\~{a}o, 65080-805, S\~{a}o Lu\'{i}s, Maranh\~{a}o, Brazil}

\affiliation{Coordena\c c\~ao do Curso de F\'{\i}sica -- Bacharelado, Universidade Federal do Maranh\~{a}o, 65085-580 S\~{a}o Lu\'{\i}s, Maranh\~{a}o, Brazil}

\begin{abstract}
We investigate the optical response of a conduction electron in a helically twisted mesoscopic medium containing a screw dislocation and a uniform torsional background, in the presence of an axial magnetic field and an Aharonov--Bohm flux. We show that the coupling between longitudinal motion and the geometric background produces an effective in-plane confinement, allowing bound states to emerge without the need for an external radial potential. Exact analytical solutions are obtained for the energy spectrum and radial wave functions, and these results are used to evaluate linear and third-order nonlinear absorption, changes in the refractive index, the photoionization cross section, and the oscillator strength. The combined action of torsion, magnetic field, and topological defect increases the interlevel spacing, compresses the radial electronic distribution, and breaks the dynamical symmetry between opposite angular-momentum channels, leading to strongly asymmetric and state-resolved optical spectra. Under intense optical excitation, the nonlinear contribution can overcome linear absorption, driving the system into a negative-absorption regime and enabling geometry-controlled optical gain. These results establish torsion and defect engineering as effective tools for tuning confinement, resonant energies, and selective amplification in mesoscopic nanophotonic platforms operating in the mid-infrared and terahertz ranges.
\end{abstract}

\maketitle

\section{Introduction}
\label{sec:intro}

Controlling light-matter interaction in low-dimensional semiconductors has become a key route toward active nanophotonic elements, including ultracompact modulators, saturable absorbers, and gain blocks operating from the terahertz to the mid-infrared \cite{RevModPhys.88.025005,PhysRevB.40.1620,PhysRevB.80.115316,PhysRevB.82.115210,PhysRevB.110.134513,LSA.2024.13.30,MRL.2023.11.21}. In nanostructures such as quantum wells, nanowires, dots (QDs), and rings (QRs), spatial confinement reshapes the density of states and enforces strict selection rules. This enables sharp, tunable optical resonances harnessed for photodetection, nonlinear optics, and on-chip photonics \cite{Levine1993QWIP,Miller1984QCSE_PRL,Miller1986QCSE_Devices,Konstantatos2010CQD_PD_Review,Rogalski2002QDIP_Review,Yan2009NanowirePhotonics,Arakawa1982QDlaser_APL,Wang2019QD_SiPhot,Lorke2000QR_PRL,Warburton2000QD_Nature}. Beyond electrostatic gating or lithographic patterning, mechanical and structural control-including built-in strain, twist, stacking angle, and dislocation engineering in van der Waals heterostructures and III-V nanostructures-now provides an additional physical parameter to tailor band offsets, excitonic resonances, and nonlinear susceptibilities \cite{YU2025130856,10.1063/5.0037852,PhysRevB.109.094105,PhysRevB.109.115434,Li2024}. In other words, geometry itself is becoming an actuator for nonlinear optical responses.

An emerging direction is to regard extended lattice defects not just as sources of scattering, but as engineerable geometric backgrounds for quantum electrons. In the geometric theory of defects, a medium containing a uniform density of parallel screw dislocations is described as a space with constant torsion and essentially vanishing curvature \cite{Katanaev1992AnnPhys,Kleinert1989GaugeFields,Nelson1987DefectsGeometry,Jensen1991,Puntigam1997CQG,Katanaev2005PhysUsp,Furtado1999PLA,Moraes2000IJMPA}. In such a background, the electron's longitudinal motion is kinematically locked to azimuthal winding: moving forward in $z$ necessarily drags a twist in $\varphi$. This coupling appears in the effective metric and, after minimal coupling to the electromagnetic field, manifests as an emergent in-plane confining potential even in the absence of a lithographically defined quantum dot. Torsion, therefore, acts like a built-in mechanical quantum well.

In mesoscopic rings and dots, this geometric confinement does not act in isolation. It coexists with two other well-established ingredients. First, a perpendicular magnetic field $\mathbf{B}$ produces cyclotron quantization and the familiar Fock-Darwin ladder of magneto-confinement levels  \cite{Landau1930,Fock1928,Darwin1930,Maksym1990PRL,McEuen1991PRL,Tarucha1996PRL,Reimann2002RMP,Beenakker1991SSP}. Second, an Aharonov-Bohm (AB) flux $\Phi$ threading the symmetry axis imprints a topological phase shift on orbiting carriers even where the local magnetic field is zero \cite{Aharonov1959PR,Aharonov1961PR,Chambers1960PRL,Tonomura1986Nature,Webb1985PRL,Washburn1992RPP,Levy1990PRL,Chandrasekhar1991PRL,Gefen1984PRL,Buttiker1983PLA,Cheung1988PRB,Lorke2000PRL}. Both ingredients reshape the allowed angular momenta, lift degeneracies, and bias the dipole matrix elements, directly modulating the optical response \cite{Buttiker1983PLA,Gefen1984PRL,Webb1985PRL,Mailly1993PRL,Lorke2000PRL,Reimann2002RMP,Viefers2004PhysE,Sheng2002PRB,Fomin2003PRB,Fomin2014Book}.

The combination of torsion, axial magnetic field, and AB flux produces three qualitative effects that, to our knowledge, have not been unified in a single microscopic platform:

\begin{enumerate}
\item \textbf{Geometric confinement and spectral tuning.} Uniform torsion acts as an effective in-plane harmonic trap whose strength scales with the torsion density and the longitudinal wave vector $k_z$. This geometric confinement blueshifts intersublevel transition energies by increasing the level spacing, in the same spirit that stronger radial confinement in a quantum dot pushes excited states to higher energies. Crucially, this effect is continuous and tunable: changing the torsion density mechanically shifts the resonance photon energy in the mid-IR/THz range without changing the material composition.

\item \textbf{Intensity-driven nonlinear switching and gain.} Because the same confined transition dominates both the linear absorption $\alpha^{(1)}(\omega)$ and its third-order (Kerr-like/saturable) correction $\alpha^{(3)}\left(\omega, I_0\right)$, a sufficiently strong drive can bleach and even invert the transition. The total absorption $\alpha\left(\omega, I_0\right)=\alpha^{(1)}+\alpha^{(3)}$ can become negative near resonance, indicating a gain-like (amplifying) response without requiring conventional population inversion between distinct bands. This defines an optical switching threshold based on intensity rather than carrier injection.

\item \textbf{State-resolved asymmetric absorption and selective gain.}  
Dipole selection rules in this cylindrical geometry enforce $\Delta m= \pm 1$. In a pristine, flux-free system, transitions from different initial states (e.g., $m=+1$ versus $m=0$ ) following the same selection rule might present similar spectral features. Here, however, the interplay of torsion, screw dislocation, and AB flux explicitly breaks the dynamical $m \leftrightarrow-m$ symmetry. As a result, the effective centrifugal barrier becomes strongly dependent on the initial orbital angular momentum. Consequently, transitions obeying the same selection rule (e.g., $\Delta m=-1$ ) occur at vastly different photon energies and exhibit entirely different oscillator strengths. Under a strong optical drive, this built-in spectral separation allows a specific transition channel to enter the negative-absorption (gain) regime, while others remain non-resonant or lossy, thereby enabling state-selective nonlinear amplification.
\end{enumerate}

All of these effects are visible in standard optical observables. Linear absorption and the real part of the refractive index directly track interlevel spacings, linewidths, and dipole overlaps. Their third-order corrections, driven by a strong pump field, encode saturation, bleaching, and the onset of a gain-like response \cite{SheikBahae1990OL,SheikBahae1991JOSAB,Wherrett1984JOSAB,ChemlaShah2001Nature,Dinu2003APL,Autere2018AdvMater,Eggleton2011NatPhoton,Kippenberg2018Science}. The photoionization cross section (PCS) provides a complementary probe of bound-tocontinuum coupling, routinely measured in similar mid-IR/THz and tesla-scale regimes \cite{Bastard1990PRB_PCS_QW,Bryant1988PRB_QD_Impurity,Sari1996PRB_QD_PCS,Ishikawa1995PRB_QW_PCS,Szeftel1992PRB_QD_Acceptor_PCS,Chakraborty1994PRB_Rings_AB_PCS,Govorov2004PRB_Rings_Exciton_AB,Maksym1990PRB_FD_Optics,Lugo2006PRB_QD_PCS_Field,Xie2013SSC_QR_PCS,Zhang2010JAP_QD_PCS_SizeField,Sadeghi2011PLA_QR_PCS_AB,Jin2018SciRep_QD_PCS_Exp,Kuyucu2014SST_QD_PCS_Review,Bai2021Nanophotonics_PCS_Topology,CTP.2024.76.105701}. In our setting, torsion blueshifts the PCS resonance while simultaneously quenching its amplitude by reducing dipole overlap, providing a purely spectroscopic fingerprint of geometric confinement.

In this work, we propose a torsion-engineered, magnetically biased mesoscopic electron system as a geometric control platform for nonlinear and state-selective nanophotonics. We consider a nonrelativistic electron of effective mass $m^*$ in a medium modeled as a uniform torsion background, simultaneously subjected to an axial magnetic field and an Aharonov-Bohm (AB) flux. In Sec.~\ref{sec:helical_metric},  starting from the minimally coupled Schrödinger equation, we derive an exact radial equation and show that torsion generates an in-plane confining potential, obtaining closed-form eigenenergies and normalized radial wave functions. In Secs.~\ref{sec:optical_properties_framework} and \ref{sec:results}, we use these analytic inputs to compute the linear absorption, third-order absorption, refractive index changes, and photoionization cross section. We demonstrate an intensitydriven inversion of the absorption peak, yielding a torsion-tunable nonlinear optical gain channel. Furthermore, we show that torsion and AB flux jointly break the $m \leftrightarrow -m$ symmetry of the dipole-allowed transitions, producing strongly state-selective gain. We argue that torsion-induced confinement provides a practical geometric parameter to spectrally target and selectively amplify specific angular momentum channels in the mid-IR/THz range, without relying on conventional population inversion.

\section{Schr\"odinger equation in a helically twisted background}
\label{sec:helical_metric}

We consider a charged particle moving in a three-dimensional medium described by the helically twisted metric \cite{SilvaNetto2008JPCM,PLA.2012.376.2838,universe.2025.11.391,AoP.2026.484.170295,PLB.2025.870.139944,PE.2026.179.116497}
\begin{equation}
 ds^2 = \bigl[dz +(\beta+ \tau\rho^2)\, d\varphi\bigr]^2 + d\rho^2 + \rho^2\, d\varphi^2,
 \label{eq:metric}
\end{equation}
where $(\rho,\varphi,z)$ are cylindrical coordinates, $\beta$ is the screw-dislocation parameter, and $\tau$ controls the uniform torsion density. The nontrivial mixing between $z$ and $\varphi$ encodes the helical distortion of the medium.

\begin{figure*}[tbhp]
    \centering
    \includegraphics[width=0.7\linewidth]{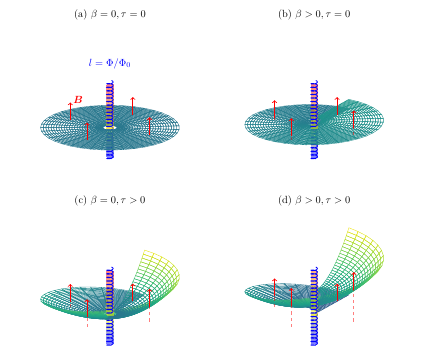}
    \caption{\footnotesize
    (Color online) Embedding surface geometry for different ($\beta, \tau$) combinations: (a) Euclidean ($\beta=0, \tau=0$); (b) pure screw dislocation ($\beta>0, \tau=0$); (c) pure torsion ($\beta=0, \tau>0$); (d) combined geometry. Red arrows indicate the magnetic field $B$; the blue solenoid indicates the Aharonov-Bohm flux $\Phi$.}
    \label{fig:torsion}
\end{figure*}

Figure \ref{fig:torsion} illustrates the geometric roles of $\beta$ and $\tau$. While the screw dislocation yields a linear vertical displacement (standard helicoid), the uniform torsion induces a non-linear winding that depends on the radial distance. The setup also includes a uniform magnetic field $\boldsymbol{B}$ and an Aharonov-Bohm flux $l$, as depicted.

\subsection{Metric tensor, its inverse, and the Laplace--Beltrami operator}

From Eq.~\eqref{eq:metric}, the spatial metric $g_{ij}$ in the coordinate basis $(\rho,\varphi,z)$ has the following nonvanishing components \cite{OQE.2026.58.113,AdP.2026.538.e00593,PE.2026.179.116497,PLB.2025.870.139944,AoP.2026.484.170295,universe.2025.11.391}:
\begin{align}
 g_{\rho\rho} &= 1, 
 \label{eq:g_rhorho}\\[2pt]
 g_{\varphi\varphi} &= \rho^2 + \bigl(\beta + \tau\rho^2\bigr)^2, 
 \label{eq:g_phiphi}\\[2pt]
 g_{\varphi z} &= g_{z\varphi} = \beta + \tau\rho^2, 
 \label{eq:g_phiz}\\[2pt]
 g_{zz} &= 1.
 \label{eq:g_zz}
\end{align}
For convenience, we introduce the shorthand notation:
\begin{equation}
  f(\rho) \equiv \beta + \tau\rho^2 .
  \label{eq:A_rho_def}
\end{equation}
In matrix form, the metric reads:
\begin{equation}
 g_{ij} =
 \begin{pmatrix}
  1 & 0 & 0 \\
  0 & \rho^2 + f^2(\rho) & f(\rho) \\
  0 & f(\rho) & 1
 \end{pmatrix}.
 \label{eq:g_matrix}
\end{equation}
The metric determinant is $g=\operatorname{det}\left(g_{i j}\right)=\rho^2$, yielding $\sqrt{g}=\rho$. The inverse metric components $g^{i j}$, defined by $g^{i k} g_{k j}=\delta^i{ }_j$, are:
\begin{align}
 g^{\rho\rho} &= 1, 
 \qquad
 g^{\rho\varphi} = g^{\rho z} = 0,
 \label{eq:g_inv_rhorho}\\[4pt]
 g^{\varphi\varphi} &= \frac{1}{\rho^2},
 \label{eq:g_inv_phiphi}\\[2pt]
 g^{\varphi z} &= g^{z\varphi} = -\,\frac{f(\rho)}{\rho^2},
 \label{eq:g_inv_phiz}\\[2pt]
 g^{zz} &= 1 + \frac{f^2(\rho)}{\rho^2}.
 \label{eq:g_inv_zz}
\end{align}

The Laplace--Beltrami operator acting on a scalar wave function is defined as:
\begin{equation}
 \nabla^{2}_{g} \psi
  = \frac{1}{\sqrt{g}}\,
    \partial_{i}\!\left(\sqrt{g}\,g^{ij}\partial_{j}\psi\right),
 \label{eq:LB_def}
\end{equation}
with $i,j=\rho,\varphi,z$. Using $\sqrt{g}=\rho$ and Eqs.~\eqref{eq:g_inv_rhorho}--\eqref{eq:g_inv_zz}, a straightforward calculation yields:
\begin{align}
 \nabla^{2}_{g}
 &= \frac{1}{\rho}\frac{\partial}{\partial \rho}
\left(\rho\,\frac{\partial}{\partial \rho}\right)
+ \frac{1}{\rho^2}\,\frac{\partial^2}{\partial\varphi^2}- \frac{2f(\rho)}{\rho^2}\,\frac{\partial^2}{\partial\varphi\,\partial z} \notag \\
&\quad + \left(1 + \frac{f^2(\rho)}{\rho^2}\right)\frac{\partial^2}{\partial z^2}
 \notag\\
 &= \frac{1}{\rho}\frac{\partial}{\partial \rho}\left(\rho\,\frac{\partial}{\partial \rho}\right)+ \frac{1}{\rho^2}\bigl(\partial_\varphi - f(\rho)\,\partial_z\bigr)^2+ \frac{\partial^2}{\partial z^2}.
 \label{eq:LB_compact}
\end{align}
Equation~\eqref{eq:LB_compact} explicitly shows that the helical geometry couples the azimuthal and longitudinal derivatives through the combination $\partial_\varphi - f(\rho)\,\partial_z$.

\subsection{Magnetic field, Aharonov--Bohm flux, and Hamiltonian}

We now introduce a uniform perpendicular magnetic field and an Aharonov-Bohm (AB) flux threading the $z$-axis. The dimensionless AB parameter is denoted by \cite{PE.2026.176.116416,PE.2023.459.169547,QR.2024.6.677,AdP.2023.535.202200371,CTP.2024.76.105701}:
\begin{equation}
    l \equiv \frac{\Phi}{\Phi_0} \equiv \frac{e\Phi}{2\pi\hbar},
\end{equation}
where $\Phi_0=h / e$ is the magnetic flux quantum. The single-particle Hamiltonian is the minimally coupled Schr\"odinger operator in this torsion-bearing background:
\begin{equation}
 H=\frac{1}{2 m^*}\left(-i\hbar\nabla - e\boldsymbol{A}\right)^2,
 \label{eq:H_maintext}
\end{equation}
where $D_i$ represents the gauge-covariant derivative associated with Eq.~\eqref{eq:metric}, and $m^*$ is the effective mass. The total vector potential $\mathbf{A}$ comprises two components: (i) a uniform perpendicular magnetic field $\mathbf{B}=B \hat{z}$, written in the symmetric gauge as $\mathbf{A}_B=(B \rho / 2) \hat{\varphi}$, and (ii) an AB flux confined to the $z$-axis, with $\mathbf{A}_{A B}= (\Phi / 2 \pi \rho) \hat{\varphi}$. Thus \cite{EPJC.2015.75.321,PLA.2016.380.3847},
\begin{equation}
 \boldsymbol{A} = \boldsymbol{A}_B + \boldsymbol{A}_{\text{AB}}
 = \left(\frac{B\rho}{2} + \frac{\Phi}{2\pi\rho}\right)\hat{\varphi},
 \label{eq:vector_potential_main}
\end{equation}
combining both magnetic quantization and a purely topological (AB) phase.

In cylindrical coordinates, the kinetic operator in Eq.~\eqref{eq:H_maintext} can be expressed as:
\begin{equation}
 H = -\frac{\hbar^2}{2m^*}\,
     \frac{1}{\sqrt{g}}\,
     D_i\!\left(\sqrt{g}\,g^{ij} D_j\right),
 \label{eq:H_covariant_LB}
\end{equation}
where $D_i=\partial_i-i \frac{e}{\hbar} A_i$, and $A_i$ are the covariant components of the vector potential. For the purely azimuthal vector potential given in Eq.~\eqref{eq:vector_potential_main}, the only nonvanishing component is $A_\varphi(\rho)$, the only nonvanishing component is $A_{\varphi}(\rho)$. Consequently, minimal coupling in the azimuthal direction effectively dictates the substitution:
\begin{equation}
 \partial_\varphi \;\longrightarrow\;
 \partial_\varphi - i\frac{e}{\hbar}\,\rho A_\varphi(\rho)
 = \partial_\varphi - i\left[
     \frac{eB}{2\hbar}\,\rho^2 + l
   \right],
 \label{eq:phi_substitution}
\end{equation}
within the $\left(1 / \rho^2\right)\left(\partial_{\varphi}-f(\rho) \partial_z\right)^2$ term in the Laplace-Beltrami operator. This serves as the curved-space analog of the standard Landau and AB substitutions found in flat cylindrical coordinates.

\subsection{Separation of variables and radial equation}

Because the metric coefficients and the vector potential are independent of $\varphi$ and $z$, the corresponding canonical momenta are conserved. Thus, the wave function can be separated as:
\begin{equation}
 \psi(\rho,\varphi,z)
 = \frac{1}{\sqrt{2\pi}}\,
   e^{i m \varphi}\,e^{i k z}\,R(\rho),
 \label{eq:separation_ansatz}
\end{equation}
where $m \in \mathbb{Z}$ is the azimuthal quantum number, $k_{z}$ is the longitudinal wave number, and $R(\rho)$ is the radial part.

Using Eqs.~\eqref{eq:LB_compact} and~\eqref{eq:phi_substitution}, the differential operator
\begin{equation}
 \mathcal{L} \equiv \frac{1}{\sqrt{g}}\,
 D_i\!\left(\sqrt{g}\,g^{ij} D_j\right)
\end{equation}
acting on the separated state~\eqref{eq:separation_ansatz} yields:
\begin{align}
 \mathcal{L}\psi
 &= \left\{
      \frac{d^2 R}{d\rho^2}
      + \frac{1}{\rho}\frac{dR}{d\rho}
      - \frac{L^2(\rho)}{\rho^2}\,R
      - k_{z}^2 R
    \right\}
    \frac{e^{i m\varphi}e^{i k z}}{\sqrt{2\pi}},
 \label{eq:L_on_psi}
\end{align}
where we have defined:
\begin{equation}
 L(\rho)
  \equiv  j - \Omega\,\rho^2,
 \label{eq:Lambda_def}
\end{equation}
and we have introduced the shorthand
\begin{align}
    j &\equiv m - 1 - k_{z}\beta,
    \label{eq:M0_def}\\[2pt]
    \omega &\equiv \frac{eB}{2\hbar} + k_{z}\tau.
    \label{eq:Gamma_def}
\end{align}

The stationary Schrödinger equation $H \psi=E \psi$ with the Hamiltonian given in Eq.~\eqref{eq:H_covariant_LB} then reduces to the radial equation:
\begin{equation}
 \frac{d^2 R}{d\rho^2}
 + \frac{1}{\rho}\frac{dR}{d\rho}
 + \left[\frac{2m^* E}{\hbar^2}- k_{z}^2
- \frac{L^2(\rho)}{\rho^2}
\right] R = 0.
 \label{eq:radial_R_rho_general_B}
\end{equation}
Substituting $L(\rho)=j - \Omega\rho^2$ from Eq.~\eqref{eq:Lambda_def}, we expand the terms:
\begin{equation}
  L^2(\rho)
  = \bigl(j - \Omega \rho^2\bigr)^2
  = j^2 - 2j\Omega\rho^2 + \Omega^2\rho^4.
 \label{eq:Lambda_square}
\end{equation}
Thus,
\begin{align}
 -\frac{L^2(\rho)}{\rho^2}
 &= -\frac{j^2}{\rho^2}
    + 2j\Omega
    - \Omega^2\rho^2.
 \label{eq:Lambda_over_rho2}
\end{align}
This allows Eq.~\eqref{eq:radial_R_rho_general_B} to be rewritten as:
\begin{equation}
 \frac{d^2 R}{d\rho^2}
 + \frac{1}{\rho}\frac{dR}{d\rho}
 + \left[
      \kappa^2
      - \Omega^2\rho^2
      - \frac{j^2}{\rho^2}
   \right] R(\rho) = 0,
 \label{eq:radial_R_rho_simplified}
\end{equation}
where we define the constant energy shift parameter:
\begin{equation}
 \kappa^2 \equiv \frac{2m^* E}{\hbar^2} - k_{z}^2 + 2j\Omega.
 \label{eq:kappa_def}
\end{equation}
Equation~\eqref{eq:radial_R_rho_simplified} already reveals the physical effects at play: an effective harmonic confinement term ( $\propto \Omega^2 \rho^2$ ) induced jointly by the uniform magnetic field and the torsion, a centrifugal barrier term ( $\propto j^2 / \rho^2$ ) that depends on the AB parameter $l$ and the screw dislocation parameter $\beta$, and a constant energy shift encoded in $\kappa^2$.

\subsection{Schrödinger form and effective potential}

It is convenient to cast Eq.~\eqref{eq:radial_R_rho_simplified} into a one-dimensional Schrödinger form by eliminating the first derivative via the transformation
\begin{equation}
 R(\rho) = \frac{u(\rho)}{\sqrt{\rho}}.
 \label{eq:u_def}
\end{equation}
A straightforward calculation yields
\begin{equation}
 \frac{d^2 R}{d\rho^2} + \frac{1}{\rho}\frac{dR}{d\rho}
 = \frac{1}{\sqrt{\rho}}\left[ \frac{d^2 u}{d\rho^2} + \frac{1}{4\rho^2}\,u(\rho) \right].
 \label{eq:R_to_u}
\end{equation}
Inserting Eqs.~\eqref{eq:u_def} and~\eqref{eq:R_to_u} into Eq.~\eqref{eq:radial_R_rho_simplified}, multiplying the whole equation by $\sqrt{\rho}$, and rearranging terms, we obtain
\begin{equation}
 \frac{d^2 u}{d\rho^2} + \left[ \kappa^2 - \Omega^2\rho^2 - \frac{j^2 - 1/4}{\rho^2} \right] u(\rho) = 0.
 \label{eq:radial_u_effective_B}
\end{equation}
This can be written in the canonical Schrödinger-like form
\begin{equation}
 -\frac{\hbar^2}{2 m^*} \frac{d^2 u}{d \rho^2} + V_{\mathrm{eff}}(\rho) u(\rho) = E u(\rho)
 \label{eq:Schr_like_final}
\end{equation}
by identifying the effective potential as
\begin{equation}
 V_{\mathrm{eff}}(\rho) = \frac{\hbar^{2}}{2m^{*}}\left[k^{2}-2j\Omega+\Omega^{2}\rho^{2}+\frac{j^{2}-1/4}{\rho^{2}}\right],
 \label{eq:Veff_pureE}
\end{equation}
where the effective parameters $j$ and $\Omega$ follow the definitions in Eqs.~\eqref{eq:M0_def} and~\eqref{eq:Gamma_def}.

Therefore, in the helically twisted background with uniform torsion, a perpendicular magnetic field, and an AB flux, the radial dynamics are governed by an effective potential composed of a geometry- and field-induced harmonic term ( $\propto \Omega^2 \rho^2$ ) and a flux- and screw-dependent inverse-square term ( $\propto j^2 / \rho^2$ ).

In a finite-size quantum wire of length $L_z$, the longitudinal wave vector is quantized as $k_z=n_z \pi / L_z$. The geometric term $k_z \tau$ appearing in the effective frequency $\Omega$ indicates that torsion couples directly to the longitudinal momentum along the $z$ axis. Consequently, radial confinement emerges strictly for carriers propagating along the axial direction.

The strength of this torsion-induced effect is explicitly modulated by the wire length. For a fixed torsion density $\tau$, shorter wires (larger $k_z$ ) experience a stronger effective potential, resulting in more pronounced radial localization. Conversely, in the long-wire limit ( $L_z \rightarrow \infty$ ), the longitudinal momentum $k_z$ vanishes, causing the geometric contribution to dissipate. In this limit, the effective frequency reduces to the purely magnetic cyclotron frequency $(e B / 2 \hbar)$, and the system loses its geometrically induced radial trapping. Thus, the torsion-induced mechanism constitutes a length-dependent phenomenon that requires a nonzero longitudinal propagation mode ( $k_z \neq 0$ ) to be activated.

\begin{figure}[tbhp]
\centering
\includegraphics[width=\linewidth]{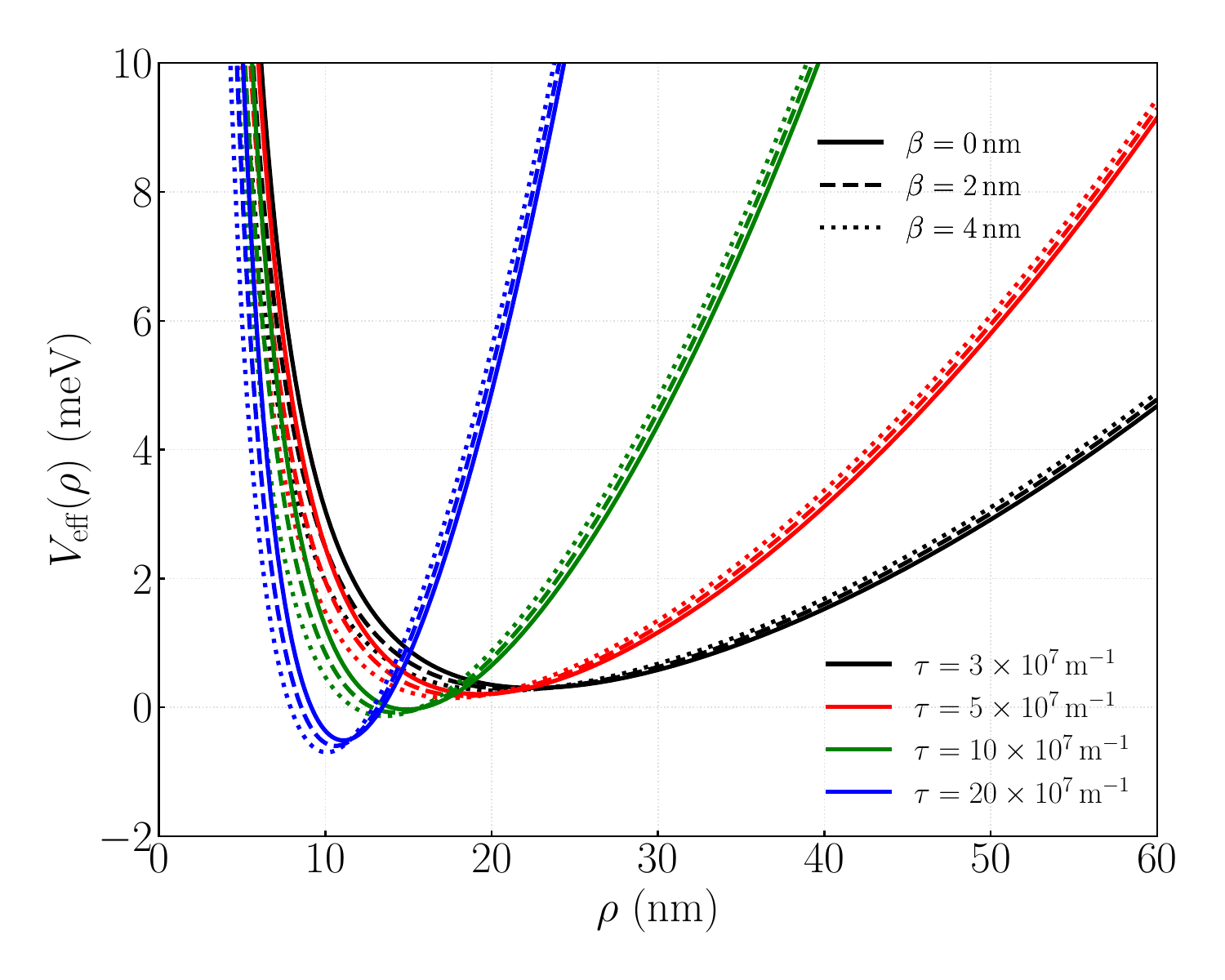}
\caption{\footnotesize
(Color online) Profile of the effective radial potential $V_{eff}(\rho)$ as a function of the radial coordinate $\rho$. Curves correspond to varying torsion densities $\tau$ (see color scale). Line styles represent different screw dislocation parameters: solid ($\beta=0$), dashed ($\beta=2$ nm), and dotted ($\beta=4$ nm).}
\label{fig:Veff}
\end{figure}

As the particle moves radially away from the axis, the quadratic term $\Omega^2 \rho^2$ in the effective potential increases, giving rise to this harmonic confinement of a purely geometric nature. Figure~\ref{fig:Veff} illustrates the dependence of the effective radial potential on the torsion density $\tau$ and the screw dislocation parameter $\beta$. It is observed that increasing $\tau$ shifts the potential minimum closer to the origin and deepens it, while simultaneously narrowing the width of the well. This behavior confirms torsion as the primary agent of confinement, capable of strongly squeezing the electron radially. Regarding the parameter $\beta$, we observe that it shifts the curve and deepens the minimum point. However, unlike torsion, the screw dislocation does not generate the harmonic trap itself; it acts only on the centrifugal barrier, thereby reconfiguring the radial equilibrium position without inducing geometric confinement.
\subsection{Bound states: radial wave functions and energy spectrum}

Equation~\eqref{eq:radial_u_effective_B} represents the radial equation of a two-dimensional isotropic oscillator with an effective frequency and a flux-shifted angular momentum. Introducing the dimensionless variable $x = |\Omega|\,\rho^{2}$ and imposing regularity at the origin along with square integrability at infinity, the solutions can be written in terms of associated Laguerre polynomials:
\begin{equation}
 u_{n,j}(\rho)
 = \mathcal{N}_{n,j}\,\rho^{|j|+\frac{1}{2}}
   \exp\!\left(-\frac{|\Omega|}{2}\rho^{2}\right)
   L_{n}^{|j|}\!\left(|\Omega|\rho^{2}\right),
 \label{eq:u_Laguerre}
\end{equation}
where $n=0,1,2,\dots$ is the radial quantum number and $j \equiv m - l - k_{z}\beta$ is the torsion- and flux-shifted angular momentum defined in Eq.~\eqref{eq:M0_def}. The original radial function $R(\rho)$ is then recovered as:
\begin{equation}
 R_{n,j}(\rho)
 = \frac{u_{n,j}(\rho)}{\sqrt{\rho}}
 = \mathcal{N}_{n,j}\,\rho^{|j|}
   \exp\!\left(-\frac{|\Omega|}{2}\rho^{2}\right)
   L_{n}^{|j|}\!\left(|\Omega|\rho^{2}\right).
 \label{eq:wavefunction_explicit}
\end{equation}

The polynomial condition on the confluent hypergeometric function yields the quantization rule
\begin{equation}
 \kappa^{2} = 2|\Omega|\left(2n + |j| + 1\right),
 \label{eq:kappa_quantization}
\end{equation}
which, upon using the definitions of $\kappa^2$, $j$, and $\Omega$, leads to the explicit energy eigenvalues:
\begin{align}
 E_{n,m}(k)
 &= \frac{\hbar}{2}
    \left|\frac{eB}{m^*} + \frac{2\hbar k_{z}\tau}{m^*}\right|
    \left(2n + \bigl|m - l - \beta k\bigr| + 1\right)
    \notag\\[2pt]
 &\quad
    - \frac{\hbar}{2}
      \left(\frac{eB}{m^*} + \frac{2\hbar k_{z}\tau}{m^*}\right)
      \bigl(m - l - \beta k\bigr)
    + \frac{\hbar^{2}k^{2}}{2m^*}.
 \label{eq:energy_eigenvalues}
\end{align}
Here, the combination
\begin{equation}
 \omega_{\text{eff}} \equiv \frac{eB}{m^*} + \frac{2\hbar k_{z}\tau}{m^*}
\end{equation}
plays the role of an effective cyclotron frequency that includes both the external magnetic field and the torsion-induced contribution. The first term in Eq.~\eqref{eq:energy_eigenvalues} has the usual Landau-like form scaling with $2n+|j|+1$, while the second term represents the AB-, screw-, and torsion-shifted ``angular drift'' proportional to $j$. The last term corresponds to the free longitudinal kinetic energy.

\begin{figure*}[htb]
    \centering
    \includegraphics[width=0.8\linewidth]{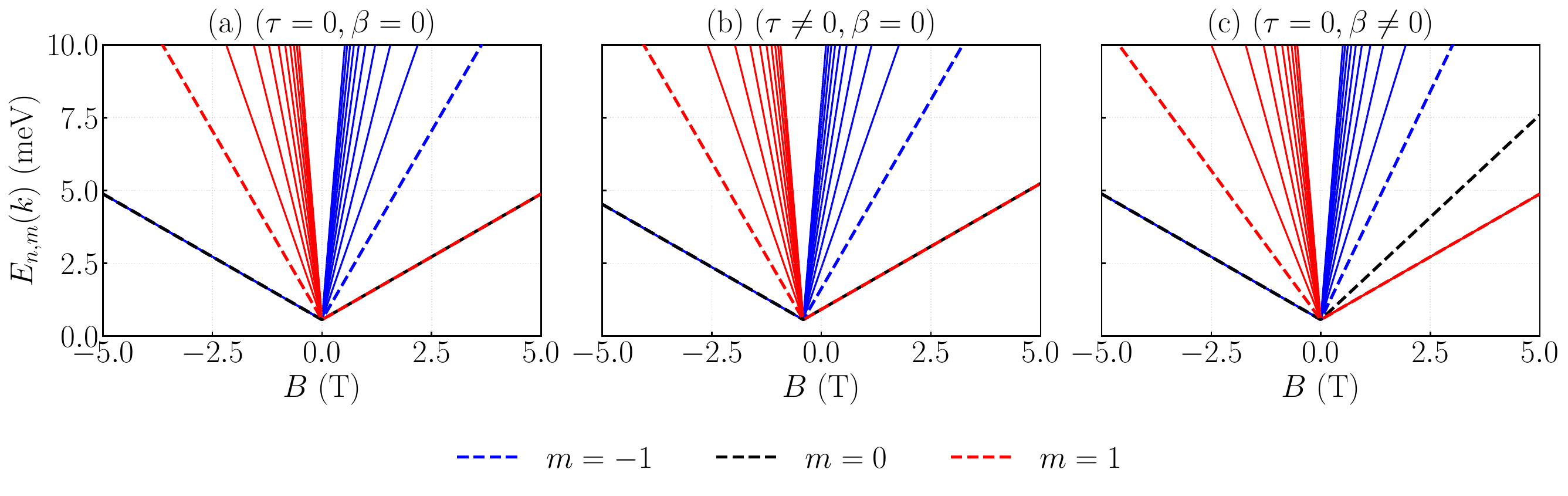}
   \caption{\footnotesize
   (Color online) Magnetic field dependence of the lowest electronic energy levels $E_{0,m}(k)$ for a quantum wire of length $L_{z}=100$ nm. (a) Standard case ($\tau=0, \beta=0$); (b) effect of torsion ($\tau\ne0$); (c) effect of screw dislocation ($\beta\ne0$).}
    \label{fig:Enm_vs_B}
\end{figure*}

Figure~\ref{fig:Enm_vs_B} illustrates the dependence of the lowest electronic energy levels, $E_{0, m}(k)$ , on the magnetic field for angular momenta in the range $m \in[-10,10]$, exploring the effects of parameters $\tau$ and $\beta$. We consider a wire of length $L_z=100 \mathrm{~nm}$ and zero Aharonov-Bohm flux $(l=0)$. Fig.~\ref{fig:Enm_vs_B}(a) presents the ideal case ( $\tau=0, \beta=$ 0 ), where the energy minimum occurs at $B=0$. For $B>0$, states with $m>0$ (red lines) and $m=0$ become degenerate, forming the Landau level, while states with $m<0$ (blue lines) undergo energy splitting, creating an asymmetric fan-like structure. The opposite behavior occurs for $B<0$. In Fig.~\ref{fig:Enm_vs_B}(b), it is noted that torsion $\tau$ shifts the spectral minimum horizontally. Finally, in Fig.~\ref{fig:Enm_vs_B}(c), the effect of $\beta$ manifests topologically, altering the band slopes and isolating the ground state $m=0$ from the Landau level degeneracy (states $m \geq 1$ in the $B>0$ region). The dashed lines highlight the states $m=-1,0,1$, which are the subject of the optical transitions studied in this work.

\begin{figure*}[tbhp]
    \centering
    \includegraphics[width=0.8\linewidth]{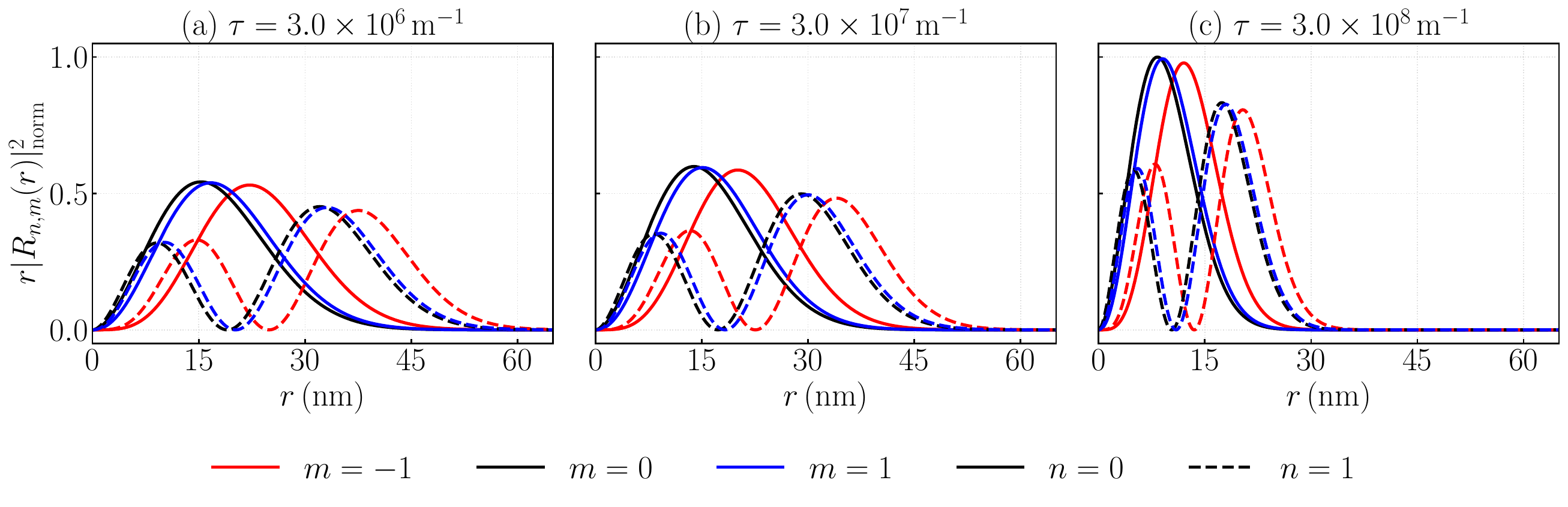}
    \caption{\footnotesize
    (Color online) Radial probability density, $r|R_{n,m}(r)|^{2}$, as a function of the radial coordinate $r$ for different torsion parameters: (a) $\tau=3.0\times10^{6}$ m$^{-1}$; (b) $\tau=3.0\times10^{7}$ m$^{-1}$; (c) $\tau=3.0\times10^{8}$ m$^{-1}$. Solid (dashed) lines correspond to the radial quantum number $n=0$ ($n=1$). Colors indicate orbital angular momenta: $m=0$ (black), $m=-1$ (red), and $m=1$ (blue). The magnetic field is fixed at $B=5$ T.}
    \label{fig:radial_density_tau}
\end{figure*}
\begin{figure*}[tbhp]
    \centering
    \includegraphics[width=0.8\linewidth]{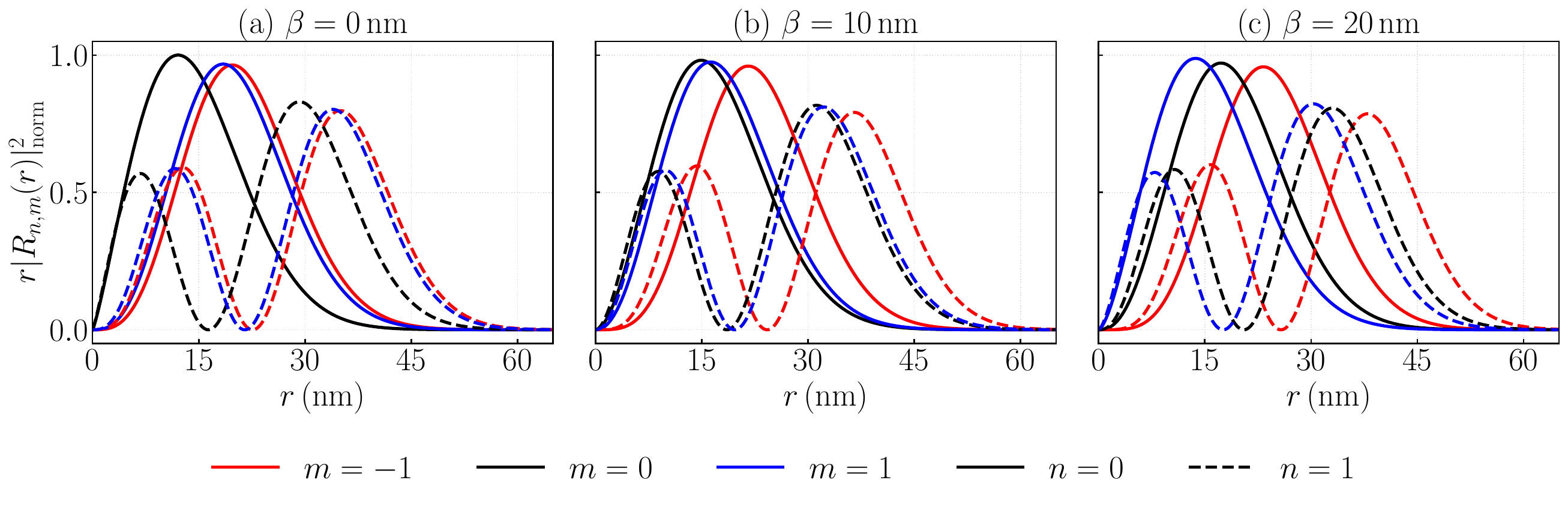}
    \caption{\footnotesize
    (Color online) Radial probability density, $r|R_{n,m}(r)|^{2}$, as a function of the radial coordinate $r$ for different screw dislocation parameters: (a) $\beta=0$ nm; (b) $\beta=10$ nm; (c) $\beta=20$ nm. Solid (dashed) lines correspond to the radial quantum number $n=0$ ($n=1$). Colors indicate orbital angular momenta: $m=0$ (black), $m=-1$ (red), and $m=1$ (blue).}
    \label{fig:radial_density_beta}
\end{figure*}
Figure~\ref{fig:radial_density_tau} presents the behavior of the radial probability density for different values of the torsion parameter $\tau$. In the low- $\tau$ regime (Fig.~\ref{fig:radial_density_tau}(a)), lateral confinement is determined primarily by the magnetic term of the effective frequency. As $\tau$ increases (Figs.~\ref{fig:radial_density_tau}(b) and (c)), he geometric contribution $\left(2 \hbar k_{z} \tau / m^*\right)$ becomes dominant in $\omega_{\text {eff }}$. Consequently, the overall increase in the effective frequency induces stronger spatial localization, squeezing the wave functions towards the origin ($r \rightarrow 0$) and highlighting the role of torsion as a geometric mechanism for controlling confinement.

In contrast, Figure~\ref{fig:radial_density_beta} investigates the role of the screw dislocation parameter $\beta$. Unlike torsion, $\beta$ does not modify the effective confinement frequency $\omega_{\text {eff }}$, and consequently does not induce wave function squeezing. Taking Fig.~\ref{fig:radial_density_beta}(a) as a reference ( $\beta=0$ ), where states with $\pm m$ exhibit identical radial distributions, the inclusion of $\beta$ acts on the effective angular momentum as a shift term, $j=m- l-\beta k_z$. This alters the centrifugal barrier $\propto j^2 / \rho^2$ in the effective potential. As evidenced in Figs.~\ref{fig:radial_density_beta}(b) and (c), increasing $\beta$ redistributes the probability weight and breaks the spatial symmetry between positive and negative $m$ states, displacing their maxima in opposite directions without compressing the overall width of the wave packet. This behavior distinguishes the topological role of the screw dislocation, which reconfigures the radial equilibrium position, from the compressive geometric confinement induced by torsion.

\section{Optical Properties: Theoretical Framework}
\label{sec:optical_properties_framework}

The geometric background defined by the uniform torsion field $\tau$ and the screw dislocation $\beta$ dictates the optical response through two complementary mechanisms. First, both parameters control the discrete spectrum $E_{n,m}$ through Eq.~\eqref{eq:energy_eigenvalues}. In particular, $\tau$ renormalizes the effective confinement frequency and typically produces a blueshift of the interlevel spacings, whereas $\beta$ shifts the effective angular momentum and lifts the symmetry between states with different $m$.

Second, the same geometric parameters reshape the radial envelopes $R_{n,m}(\rho)$ in Eq.~\eqref{eq:wavefunction_explicit}, thereby modifying the dipole matrix elements that govern the strength of the optical transitions. In this sense, $\tau$ acts mainly as a geometric squeezing mechanism, while $\beta$ primarily redistributes the radial weight by shifting the effective centrifugal barrier.

A crucial point is that the optical channels must be identified in two steps. The dipole operator first imposes the selection rule $\Delta m=\pm1$, which only determines which pairs of states can be connected by light. The actual direction of an absorption process, however, is fixed only after comparing the corresponding energies $E_{n,m}$. Therefore, for a given set of parameters $(B, \tau, \beta, l, k_z)$, the initial state of an absorption transition is always the lower-energy state of a dipole-allowed pair, while the final state is the higher-energy one. Since the spectrum in Eq.~\eqref{eq:energy_eigenvalues} is parameter-dependent, the energetic ordering of two allowed states may change, requiring the absorption arrow to be reversed accordingly. This convention is essential to consistently interpret the resonances discussed below.

\subsection{Linear and nonlinear absorption and refraction}
\label{sec:optical_properties}

We consider the system driven by a monochromatic classical field,
\begin{equation}
 \mathbf{E}(t) = E_0\, \hat{\mathbf e}_r \cos(\omega t),
 \label{eq:driving_field}
\end{equation}
which is radially polarized in the plane. Because the eigenstates are separable as $\Psi_{n,m,k_z}(\rho,\varphi,z)=R_{n,m}(\rho)e^{im\varphi}e^{ik_z z}$ [Eq.~\eqref{eq:separation_ansatz}], the in-plane dipole operator $\hat{\mathbf r}$ connects only states satisfying the selection rule
\begin{equation}
 \Delta m \equiv m_f-m_i=\pm1.
 \label{eq:dipole_selection_main}
\end{equation}
Accordingly, for any dipole-allowed pair, we define the transition matrix element as
\begin{equation}
 M_{if}\equiv \bra{\psi_f}\,\hat{\mathbf r}\,\ket{\psi_i},
 \label{eq:dipole_Mif_main}
\end{equation}
with $\psi_i\equiv\Psi_{n_i,m_i,k_z}$ and $\psi_f\equiv\Psi_{n_f,m_f,k_z}$. This same $M_{if}$ controls both absorption and refraction. The diagonal (Stark-like) terms,
$M_{ii}\equiv\bra{\psi_i}\hat{\mathbf r}\ket{\psi_i}$ and
$M_{ff}\equiv\bra{\psi_f}\hat{\mathbf r}\ket{\psi_f}$,
enter only in the nonlinear corrections.

It is important to emphasize that Eq.~\eqref{eq:dipole_selection_main} does not by itself determine the direction of the optical process. It only identifies the allowed pair. The actual absorption channel must be defined after ordering the energies from Eq.~\eqref{eq:energy_eigenvalues}. For a given dipole-allowed pair, we therefore label by $\ket{\psi_i}$ the lower-energy state and by $\ket{\psi_f}$ the higher-energy state, so that
\begin{equation}
 \Delta E\equiv E_f-E_i>0.
 \label{eq:DeltaE_main}
\end{equation}
With this convention, the absorption resonance occurs at $\hbar\omega\simeq\Delta E$, while the reverse process corresponds to emission or stimulated emission. For example, if for some parameter set one finds $E_{0,1}<E_{0,0}$, then the allowed absorption process is $\ket{0,1}\to\ket{0,0}$, even though the same pair is still governed by the selection rule $\Delta m=-1$. Thus, geometry and topology may reorder the levels, but they do not alter the dipole rule itself.

We model the optical response in terms of an effective two-level transition $|\psi_{i}\rangle\leftrightarrow|\psi_{f}\rangle$ with transition energy $\Delta E = E_f - E_i > 0$, using the eigenvalues from Eq.~\eqref{eq:energy_eigenvalues}  and a phenomenological homogeneous broadening $\Gamma$ that accounts for relaxation and dephasing. The incident intensity is denoted by $I_0$.

The total absorption coefficient is written as the sum of a linear term and a third-order nonlinear contribution,
\begin{equation}
 \alpha(\omega,I_0)
 =
 \alpha^{(1)}(\omega)
 +
 \alpha^{(3)}(\omega,I_0),
 \label{eq:alpha_total_main}
\end{equation}
where the linear part is
\begin{equation}
 \alpha^{(1)}(\omega)
 =
 \omega
 \sqrt{\frac{\mu}{\epsilon_r}}\;
 \frac{
 \sigma_s\, |M_{if}|^2\, \hbar \Gamma
 }{
 (\Delta E - \hbar \omega)^2 + (\hbar \Gamma)^2
 }.
 \label{eq:alpha_linear_main}
\end{equation}
Equation~\eqref{eq:alpha_linear_main} has the usual Lorentzian structure centered at $\hbar\omega\simeq\Delta E$, with peak amplitude proportional to the oscillator strength $|M_{if}|^2$. In the present system, increasing torsion generally increases the level spacing while simultaneously squeezing the radial wavefunctions, thereby reducing the dipole overlap and, therefore, suppressing the linear absorption amplitude.

Under stronger driving, the same transition acquires a third-order correction,
\begin{align}
 \alpha^{(3)}(\omega, I_0)
 &= - \omega
 \sqrt{\frac{\mu}{\epsilon_r}}\;
 \left(
 \frac{I_0}{2 n_r \epsilon_0 c}
 \right)
 \sigma_s
 \nonumber\\
 &\quad\times
 \frac{|M_{if}|^2\, \hbar \Gamma}{
 \big[(\Delta E - \hbar\omega)^2 + (\hbar \Gamma)^2\big]^2}
 \notag\\
 &\quad\times
 \left[
 4|M_{if}|^2
 -
 |M_{ff} - M_{ii}|^2\,\Xi(\omega)
 \right],
 \label{eq:alpha_cubic_main}
\end{align}
where
\begin{equation}
 \Xi(\omega)
 =
 \frac{
 3\Delta E^{\,2}
 - 4\Delta E\,\hbar \omega
 + \hbar^2(\omega^2 - \Gamma^2)
 }{
 \Delta E^{\,2} + (\hbar \Gamma)^2
 }.
 \label{eq:Xi_main}
\end{equation}
Near resonance, $\alpha^{(3)}$ is typically negative and grows in magnitude with $I_0$, producing the usual saturable-absorption behavior. When $|\alpha^{(3)}|>\alpha^{(1)}$, the total coefficient becomes negative, $\alpha(\omega,I_0)<0$, signaling a gain-like regime. In this picture, the negative absorption does not alter the level ordering itself; rather, it reflects the nonlinear optical response associated with the same dipole-allowed pair defined above.

The same microscopic ingredients determine the refractive response. We write the relative refractive-index change as
\begin{equation}
 \frac{\Delta n(\omega)}{n_r}
 =
 \frac{\Delta n^{(1)}(\omega)}{n_r}
 +
 \frac{\Delta n^{(3)}(\omega,I_0)}{n_r},
 \label{eq:dn_total_main}
\end{equation}
with linear contribution
\begin{equation}
 \frac{\Delta n^{(1)}(\omega)}{n_r}
 =
 \frac{\sigma_s |M_{if}|^2}{2 n_r^2 \epsilon_0}
 \,
 \frac{
 \Delta E - \hbar\omega
 }{
 (\Delta E - \hbar\omega)^2 + (\hbar \Gamma)^2
 },
 \label{eq:dn_linear_main}
\end{equation}
which exhibits the usual odd dispersive profile around resonance.

The third-order correction to the refractive index is
\begin{align}
 &\frac{\Delta n^{(3)}(\omega, I_0)}{n_r}
 =
 -\frac{\mu\, c\, \sigma_s\, I_0}{4 n_r^3 \epsilon_0}\;
 \frac{|M_{if}|^2}{
 \big[(\Delta E - \hbar\omega)^2 + (\hbar \Gamma)^2 \big]^2}
 \nonumber\\
 &\quad\times
 \bigg\{
 4(\Delta E - \hbar\omega)\,|M_{if}|^2
 -
 |M_{ff} - M_{ii}|^2\,G(\omega)
 \bigg\},
 \label{eq:dn_cubic_main}
\end{align}
where $G(\omega)$ is a detuning-dependent factor involving the same scales $\Delta E$ and $\Gamma$. As in the absorption coefficient, the cubic term distorts and amplifies the line shape in an intensity-dependent manner.

Therefore, in Eqs.~\eqref{eq:alpha_linear_main}--\eqref{eq:dn_cubic_main}, all geometric and topological effects enter only through two microscopic quantities: the transition energy $\Delta E$, obtained from the ordered pair of levels in Eq.~\eqref{eq:energy_eigenvalues}, and the corresponding dipole matrix elements built from Eq.~\eqref{eq:wavefunction_explicit}. This separation is useful because it makes clear that the selection rule identifies the allowed pair, whereas the spectrum determines which member of that pair is the initial state of absorption for a given set of parameters.

\subsection{Photoionization cross section}
\label{sec:theory_photoionization}

In addition to bound-bound absorption, we also evaluate the photoionization cross section (PCS), which measures the probability that a photon of energy $\hbar\omega$ promotes an electron from a localized state to an extended state. This quantity provides a complementary spectroscopic probe of both the transition energy and the dipole strength.

Within the electric-dipole approximation, the PCS for an initial bound state $\ket{\psi_i}$ is
\begin{align}
\sigma(\hbar\omega)
&=
\left(
\frac{n_r}{\kappa}
\right)
\frac{4\pi^2}{3}\,
\beta_{FS}\,
\hbar\omega
\sum_f
\big|
\bra{\psi_i} \hat{\mathbf r} \ket{\psi_f}
\big|^2
\nonumber\\
&\quad\times
\delta(E_f-E_i-\hbar\omega),
\label{eq:sigma_general_main}
\end{align}
where $\beta_{FS}=e^2/\hbar c$ is the fine-structure constant, $\kappa$ is the dielectric constant of the host material, and the sum runs over all dipole-allowed final states. The Dirac delta enforces energy conservation.

As before, the direction of the optical process is fixed by the energy ordering: the initial state must be the lower-energy bound state, and the final state must satisfy $E_f>E_i$. In practice, we include a finite linewidth by replacing the delta function with a Lorentzian of width $\hbar\Gamma$,
\begin{equation}
 \delta(E_f - E_i - \hbar\omega)
 \;\longrightarrow\;
 \frac{1}{\pi}
 \frac{\hbar\Gamma}{
 (E_f - E_i - \hbar\omega)^2 + (\hbar\Gamma)^2
 }.
 \label{eq:lorentzian_main}
\end{equation}

For the numerical results discussed in Sec.~\ref{sec:results}, we focus on the specific dipole-allowed pairs satisfying the condition $\Delta m = -1$, specifically involving the lowest radial mode states $|+1\rangle \rightarrow |0\rangle$ and $|0\rangle \rightarrow |-1\rangle$. The corresponding absorption channel is defined, once again, only after comparing the respective energies from Eq.~\eqref{eq:energy_eigenvalues}. In the parameter regime adopted below, these pairs are ordered such that the lower-energy state acts as the initial state and the higher-energy state as the final one. Therefore, the working PCS expression can be written as:
\begin{align}
\sigma(\hbar\omega)
&=
\left(
\frac{n_r}{\kappa}
\right)
\frac{4\pi}{3}\,
\beta_{FS}\,
\hbar\omega\,
\big|
\bra{\Psi_f}
 \hat{\mathbf r}
\ket{\Psi_i}
\big|^2
\nonumber\\
&\quad\times
\frac{\hbar\Gamma}{
(E_f - E_i - \hbar\omega)^2 + (\hbar\Gamma)^2},
\label{eq:sigma_working_main}
\end{align}
where $\ket{\Psi_i}$ and $\ket{\Psi_f}$ denote the lower- and higher-energy members, respectively, of the chosen dipole-allowed pair.

\medskip
In summary, Eqs.~\eqref{eq:alpha_total_main}--\eqref{eq:sigma_working_main} are the only optical formulas fed with microscopic input. Everything that depends on torsion ($\tau$), screw dislocation ($\beta$), magnetic field ($B$), AB flux ($l$), and longitudinal momentum ($k_{z}$) enters through:
(i) the ordered level spacing $\Delta E=E_f-E_i$ obtained from Eq.~\eqref{eq:energy_eigenvalues}; and
(ii) the dipole overlaps built from the radial envelopes $R_{n,m}(\rho)$ in Eq.~\eqref{eq:wavefunction_explicit}.
The remainder of the paper analyzes how these two quantities reshape the absorption, refraction, nonlinear switching, and photoionization spectra.

\paragraph*{Modeling note.}
Throughout, each state-resolved resonance is described by an effective two-level (single-oscillator) model. This approximation is justified whenever the spacing to neighboring dipole-allowed lines exceeds both the homogeneous width $\Gamma$ and the drive-induced Rabi scale. Within this regime, the model captures the torsion-driven blueshift, the suppression of oscillator strength, and the onset of negative absorption, while multilevel corrections become relevant only near accidental degeneracies or under stronger driving conditions.

\section{Results and Discussion}
\label{sec:results}

In this section, we analyze the effects of torsion density $(\tau)$ and screw dislocation $(\beta)$ on the optical properties of the system, specifically the absorption coefficient and the change in refractive index. Since all optical processes considered here involve only the lowest radial mode  $(n=0)$, we hereafter abbreviate the states $|n, m\rangle$ as $|m\rangle$. The relevant optical channels are first identified by the dipole selection rule $\Delta m=-1$, which singles out the pairs $(|+1\rangle,|0\rangle)$ and $(|0\rangle,|-1\rangle)$. The actual direction of each absorption process is then determined by the ordering of the energies $E_{0, m}$ for the chosen values of $B, \tau, \beta, l$ and $L_z$. In the numerical regime adopted throughout this section, the spectrum satisfies $E_{0,+1}<E_{0,0}<E_{0,-1}$, so that the two absorption channels are the low-energy transition $|+1\rangle \rightarrow|0\rangle$ and the high-energy transition $|0\rangle \rightarrow|-1\rangle$. For other parameter sets, this ordering may change; in that case, the absorption arrow must be reversed accordingly, while the selection rule itself remains unchanged.

To explore the nonlinear response, we consider four incident-intensity regimes, namely very low ($1.0 \times 10^5~\mathrm{W/m}^2$), low ($5.0 \times 10^5~\mathrm{W/m}^2$), intermediate ($16 \times 10^5~\mathrm{W/m}^2$), and high ($22 \times 10^5~\mathrm{W/m}^2$). Unless otherwise stated, when specifically examining the effects of $\beta$ and $\tau$ we fix the intensity at the high-intensity value $I_0=22 \times 10^5~\mathrm{W/m}^2$.

The optical properties were calculated numerically using typical parameters for GaAs heterostructures. The specific values of the material constants employed are summarized in Table \ref{tab:parameters}.
\begin{table}[htb]
\centering
\caption{Material and system parameters used in numerical calculations (GaAs-based system).}
\label{tab:parameters}
\begin{tabular}{lcc}
\toprule
Parameter & Symbol & Value \\
Effective mass & $m^*$ & $0.067\,m_e$ \\
Refractive index & $n_r$ & 3.2 \\
Relative permittivity & $\epsilon_r$ & 10.24 \\
Magnetic field & $B$ & 5 T \\
Wire length & $L_z$ & 100 nm \\
Damping rate & $\Gamma$ & $0.25$ meV/$\hbar$ \\
Electronic density & $\sigma$ & $1.0 \times 10^{20}$ m$^{-3}$ \\
AB parameter & $\ell$ & 0.1 \\
\end{tabular}
\end{table}

\begin{figure*}[tbhp]
    \centering
    \includegraphics[width=0.48\linewidth]{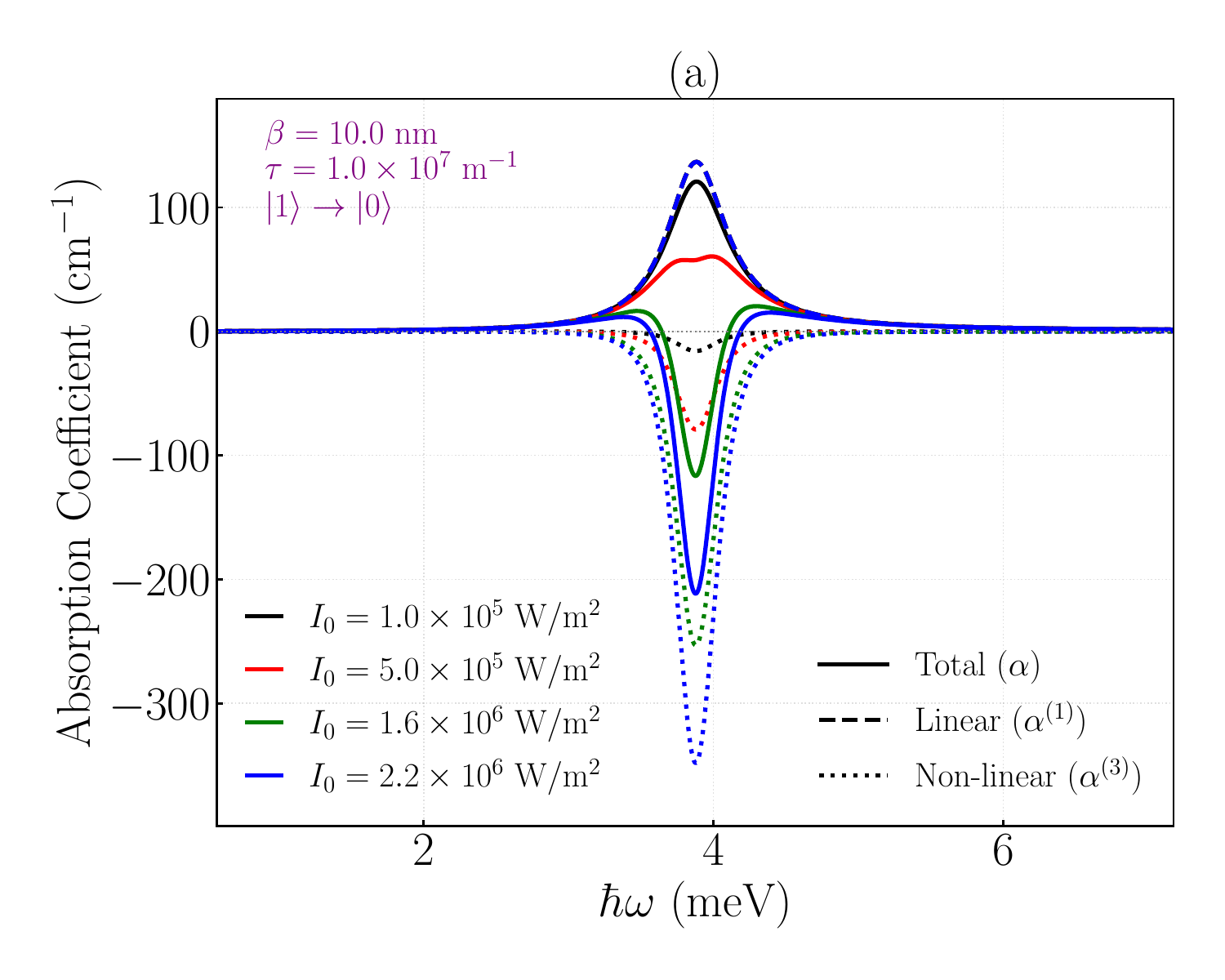}
    \includegraphics[width=0.48\linewidth]{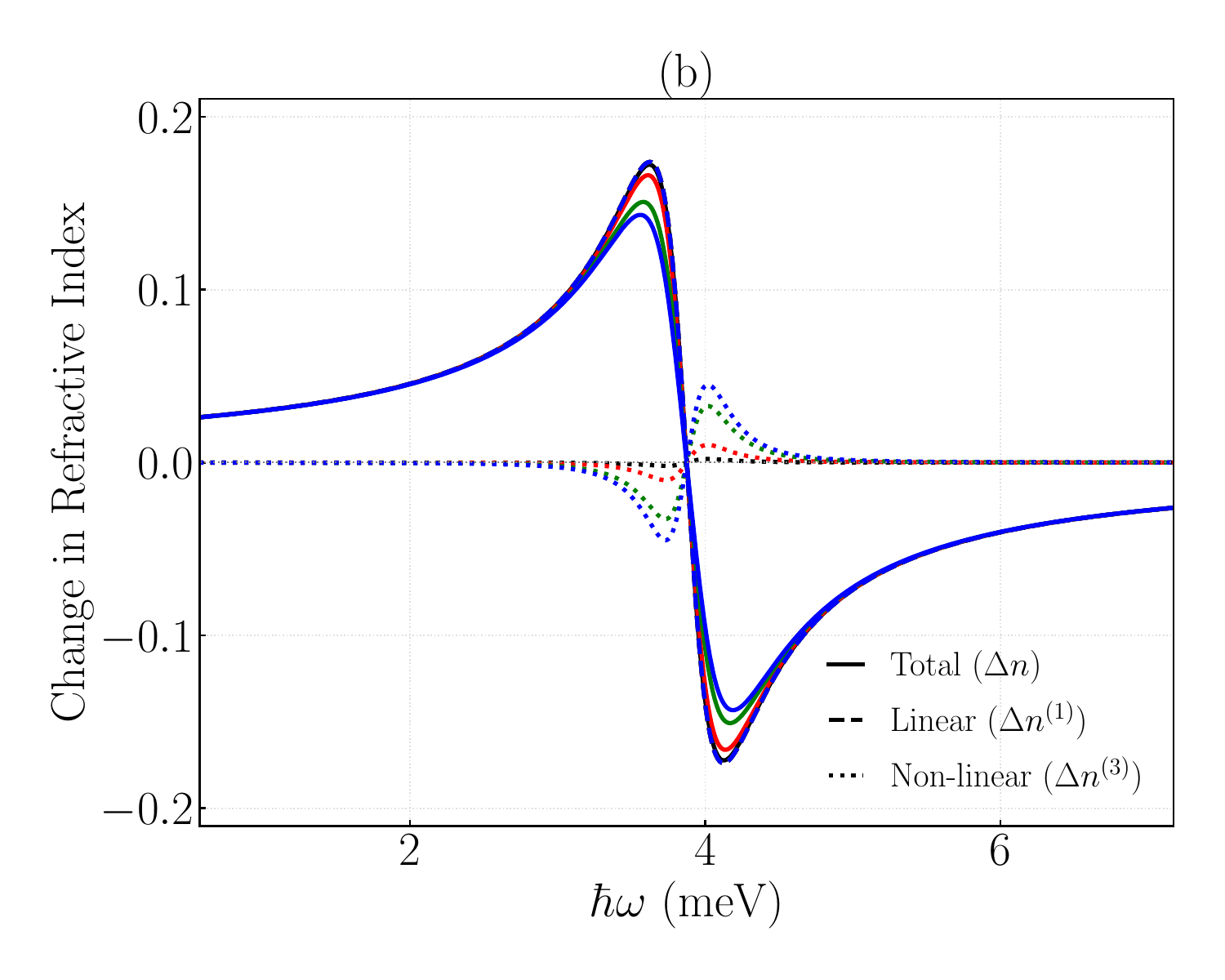}
    \caption{\footnotesize
    (Color online) Optical response as a function of the incident light intensity $I_{0}$ for the low-energy $|+1\rangle \rightarrow |0\rangle$ intersubband transition ($\Delta m=-1$). (a) Absorption coefficient $\alpha(\omega)$ and (b) refractive index change $\Delta n(\omega)/n_{r}$. Fixed parameters: $\beta=10.0$ nm and $\tau=1.0\times10^{7}$ m$^{-1}$.}
    \label{fig:I0m1}
\end{figure*}

Figure~\ref{fig:I0m1}(a) illustrates the evolution of the linear, third-order nonlinear, and total optical absorption coefficients (OACs) as a function of the incident photon energy for the low-energy absorption channel $|+1\rangle \rightarrow|0\rangle(\Delta m=-1)$, with the geometrical parameters fixed at $\beta=10 \mathrm{~nm}$ and $\tau=10^7 \mathrm{~m}^{-1}$. In the low-intensity regime ( $I_0=1.0 \times 10^5 \mathrm{~W} / \mathrm{m}^2$ ), the system exhibits a linear absorption peak with a characteristic Lorentzian profile. As the intensity increases to $2.2 \times 10^6 \mathrm{~W} / \mathrm{m}^2$, the third-order nonlinear contribution $\alpha^{(3)}$, which has an opposite sign to the linear term, becomes dominant. This behavior initially leads to absorption saturation and subsequently drives the total absorption coefficient to negative values, signaling the transition to an intense-field-induced optical gain regime.

In parallel, Fig.~\ref{fig:I0m1}(b) presents the corresponding variations of the total refractive index change (RIC) and its linear and nonlinear components. It is observed that, as the intensity ${I}_0$ increases, the overall amplitude of the refractive index variation decreases. This effect arises from the competition between the positive linear response and the negative third-order nonlinear contribution, which suppresses the refractive index modulation before inducing an inversion of the dispersion slope, a feature directly associated with the optical gain regime.

\begin{figure*}[tbhp]
    \centering
    \includegraphics[width=0.48\linewidth]{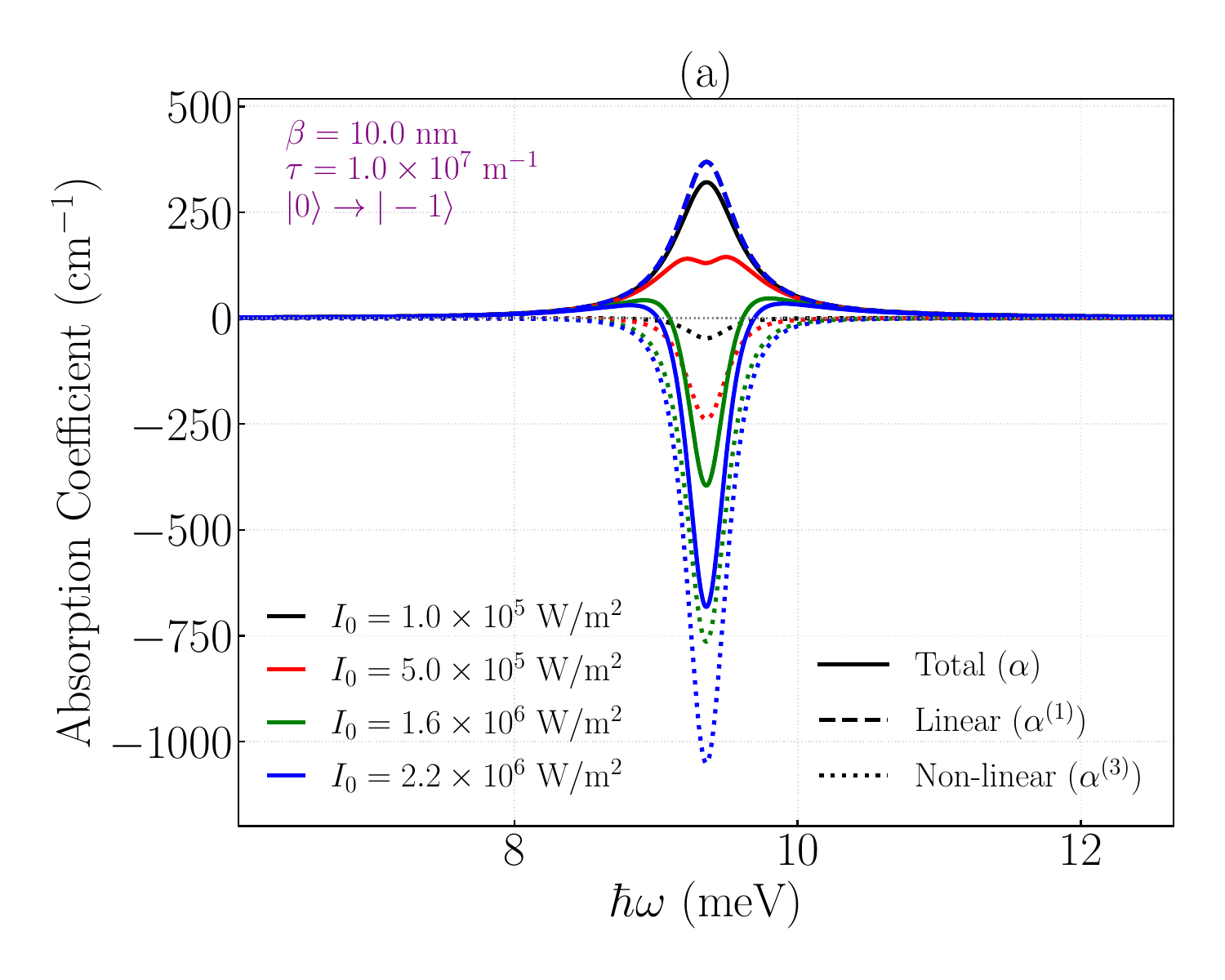}
    \includegraphics[width=0.48\linewidth]{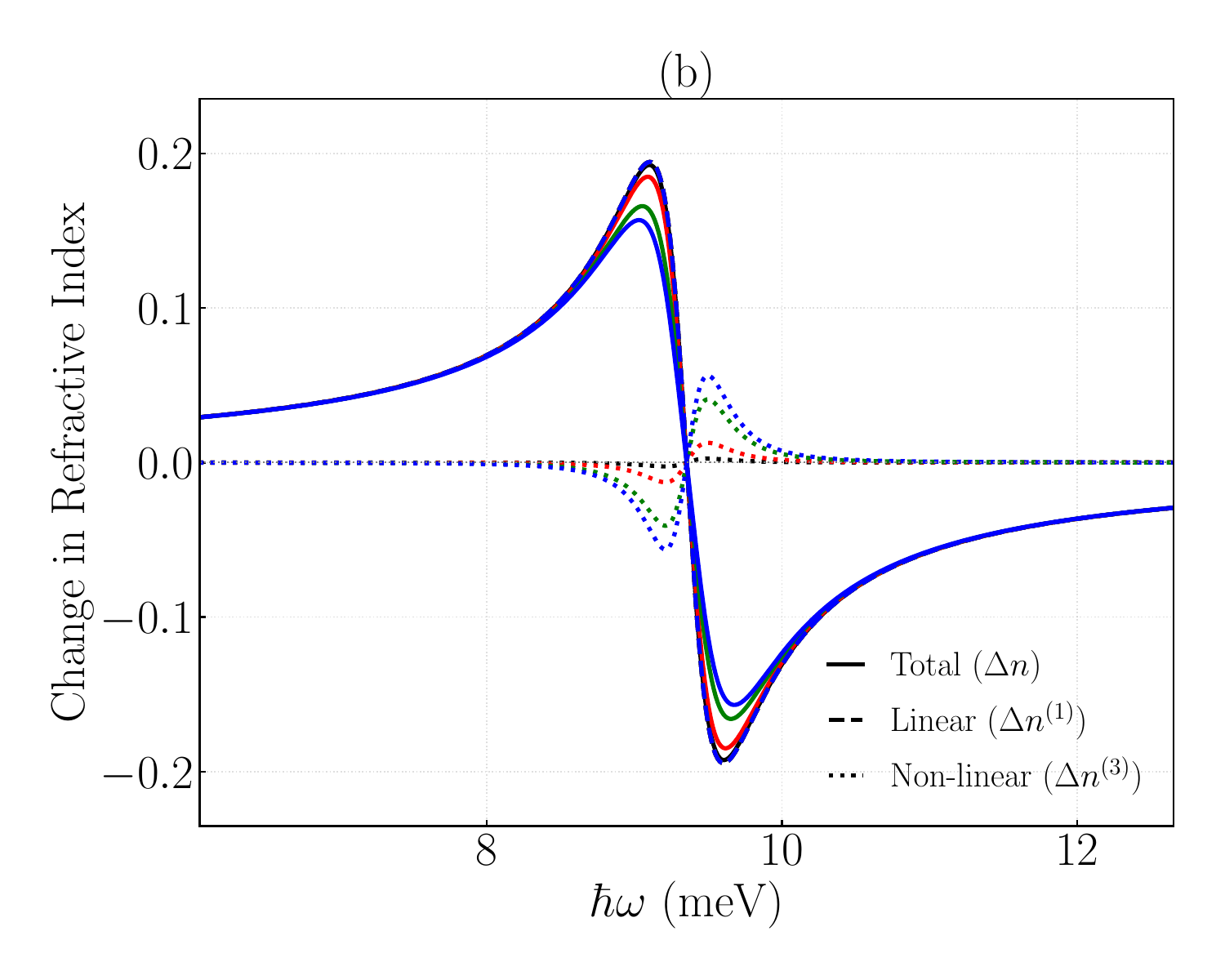}
    \caption{\footnotesize
    (Color online) Optical response as a function of the incident light intensity $I_{0}$ for the high-energy $|0\rangle \rightarrow |-1\rangle$ intersubband transition ($\Delta m=-1$). (a) Absorption coefficient $\alpha(\omega)$ and (b) refractive index change $\Delta n(\omega)/n_{r}$. Fixed parameters: $\beta=10.0$ nm and $\tau=1.0\times10^{7}$ m$^{-1}$.}
    \label{fig:I0m-1}
\end{figure*}

Figure~\ref{fig:I0m-1} displays the profiles of the optical absorption coefficients (OACs) and refractive index changes (RICs), calculated using the same parameters as in Figure~\ref{fig:I0m1}, but specifically for the high-energy absorption channel $|0\rangle \rightarrow|-1\rangle(\Delta m=-1)$.

A comparison of the absorption profiles of the two distinct $\Delta m=-1$ absorption channels reveals a pronounced spectral asymmetry, highlighting the state-dependent nature of the system under the combined action of the magnetic field and geometric parameters.

As observed in the figures, the resonance corresponding to the high-energy $|0\rangle \rightarrow |-1\rangle$ transition occurs at approximately $9.35$ meV, whereas for the low-energy $|+1\rangle \rightarrow |0\rangle$ transition, the peak shifts to a significantly lower energy, around $3.87$ meV. This energetic separation is a direct consequence of the dependence of the total energy on the effective angular momentum $j= m-l-k \beta$.

For the high-energy channel (where the final state is $m=-1$), the terms associated with the Aharonov-Bohm flux, the screw dislocation $(\beta)$, and the cyclotron frequency contribute constructively to the effective potential. This results in a tighter radial confinement and, consequently, a larger separation between energy levels, characterized by a \textit{blueshift} of the resonance.

In contrast, for the low-energy channel ($|+1\rangle \rightarrow |0\rangle$), the positive orbital angular momentum of the initial state competes with the terms associated with the geometric parameters and the magnetic flux. This leads to a partial cancellation that reduces the effective confinement frequency and shrinks the energy gap. This difference in resonance energies renders the quantum wire a highly selective medium for the electron's initial angular momentum state, thereby characterizing state-resolved asymmetric absorption behavior.

\subsection{Topological Control via Screw Dislocation (\texorpdfstring{$\beta$}{beta})}
\label{sec:topological_beta}

While torsion acts as a confinement agent by modifying the effective cyclotron frequency, the screw dislocation parameter $\beta$ plays a fundamentally different, topological role. It enters the Hamiltonian solely through the effective angular momentum $j = m - l - \beta k_z$. This modification shifts the centrifugal barrier $V_{\text{cent}} \propto j^2/\rho^2$ without altering the harmonic confinement scale $\Omega$. Consequently, $\beta$ effectively mimics the presence of an additional Aharonov-Bohm flux, allowing for fine spectral tuning and symmetry breaking.

\begin{figure*}[tbhp]
    \centering
    \includegraphics[width=0.48\linewidth]{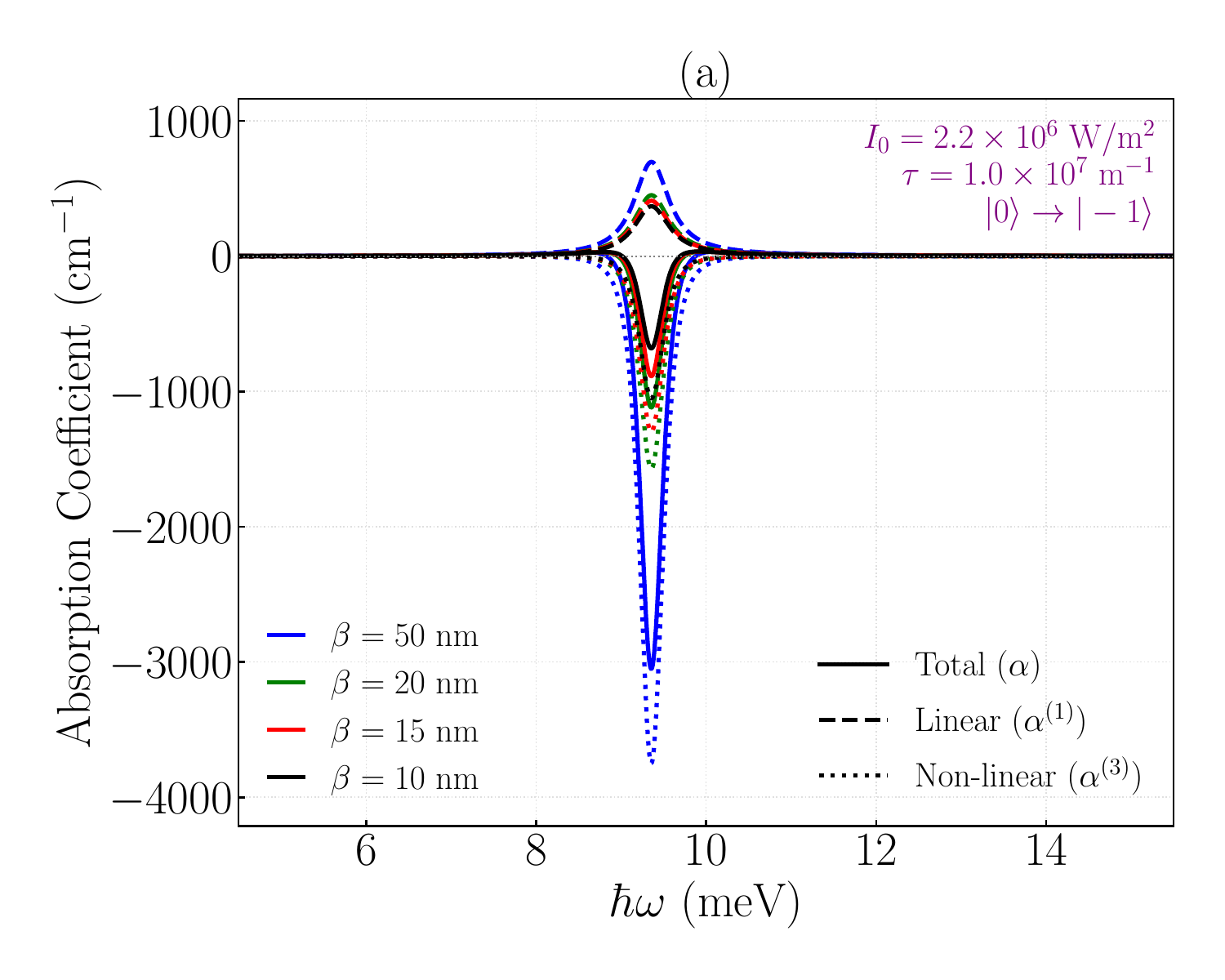}
    \includegraphics[width=0.48\linewidth]{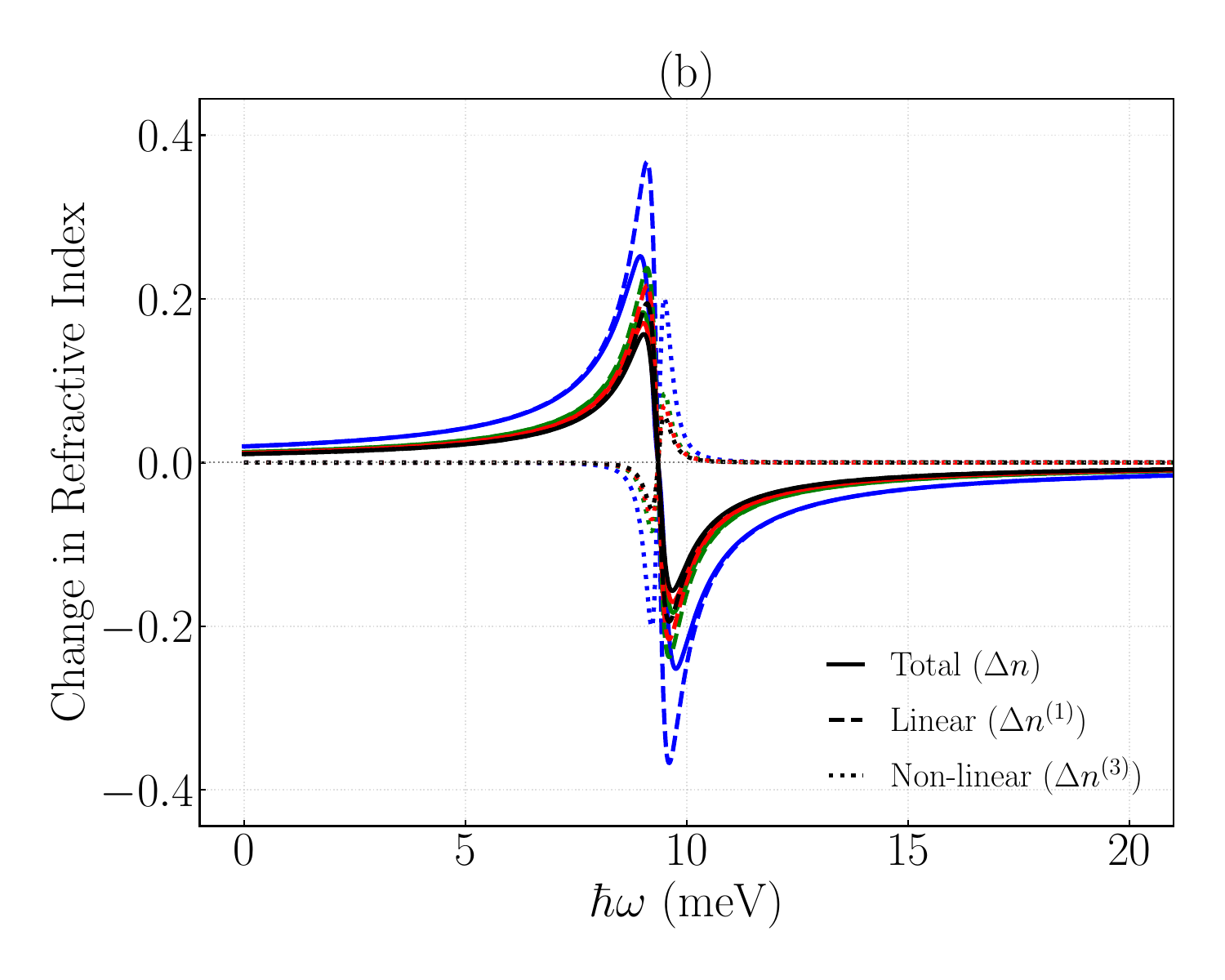}
    \caption{\footnotesize
    (Color online) Topological tuning of the high-energy $\Delta m=-1$ transition ($|0\rangle \rightarrow |-1\rangle$). (a) Absorption coefficient $\alpha(\omega)$ and (b) refractive index change $\Delta n(\omega)/n_{r}$ for different screw dislocation parameters $\beta$. Solid, dashed, and dotted lines indicate the total, linear, and third-order contributions, respectively. Fixed parameters: $I_{0}=2.2\times10^{6}$ W/m$^{2}$ and $\tau=1.0\times10^{7}$ m$^{-1}$.}
    
    \label{fig:beta_channel_neg}
\end{figure*}

In Figure~\ref{fig:beta_channel_neg}, we present the OACs and RICs for the high-energy absorption channel $|0\rangle \rightarrow |-1\rangle$. For the intensity considered ($I_0 = 22 \times 10^5$ W/m$^2$), the nonlinear contribution $\alpha^{(3)}$ dominates, driving the total absorption to negative values at resonance, a signature of geometrically induced optical gain.
As $\beta$ increases, the peak amplitude is progressively suppressed. This reflects the reduction of the dipole matrix element $|M_{if}|^2$ caused by the reinforcement of the centrifugal barrier, which displaces the radial equilibrium position and reduces the overlap between the initial and final state envelopes. 
Notably, the spectral position of the resonance remains robust. This indicates that for the $|0\rangle \rightarrow |-1\rangle$ channel, the screw dislocation acts primarily as an intensity modulator, allowing control over gain efficiency without altering the operating frequency.
\begin{figure*}[tbhp]
    \centering
    \includegraphics[width=0.48\linewidth]{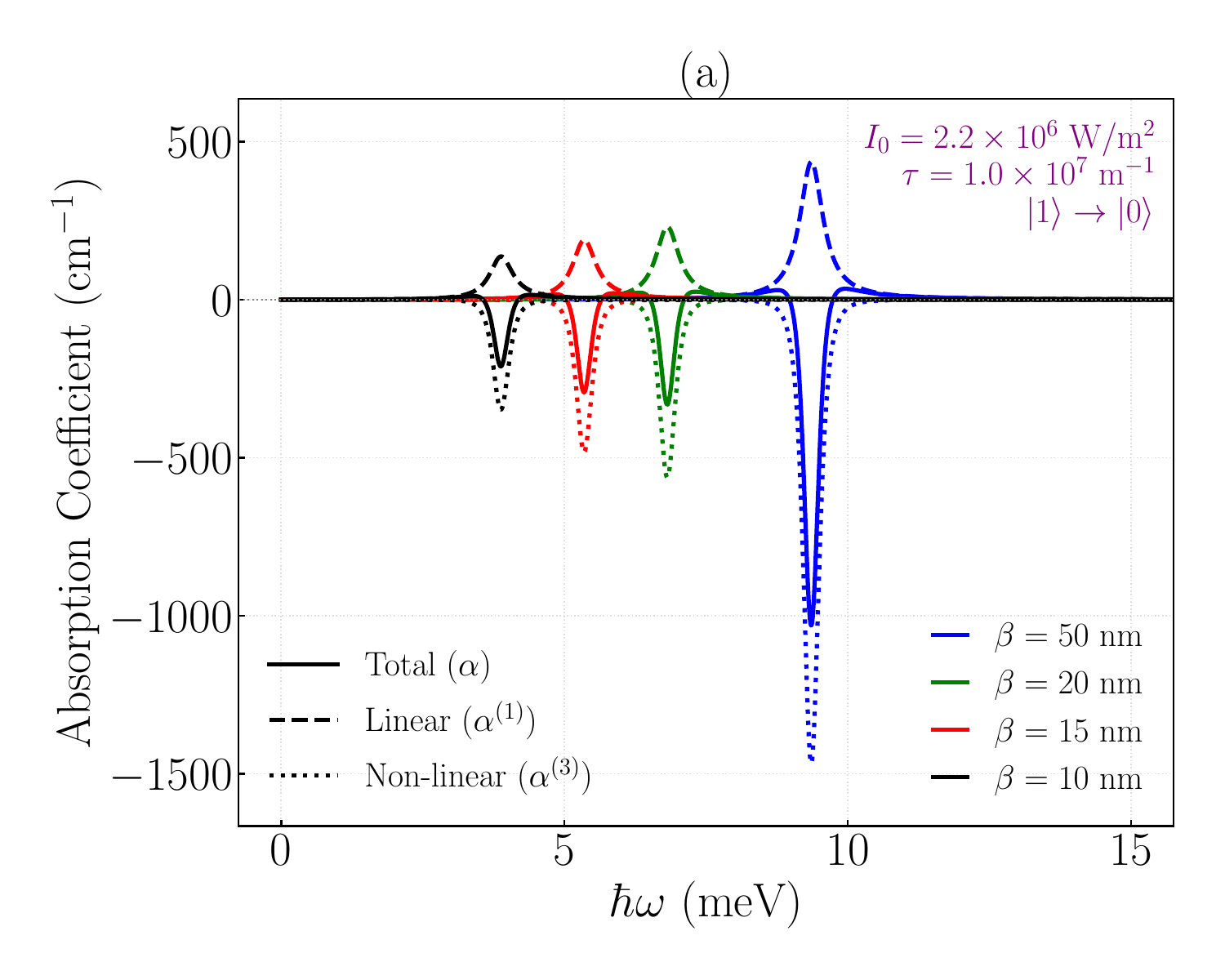}
    \includegraphics[width=0.48\linewidth]{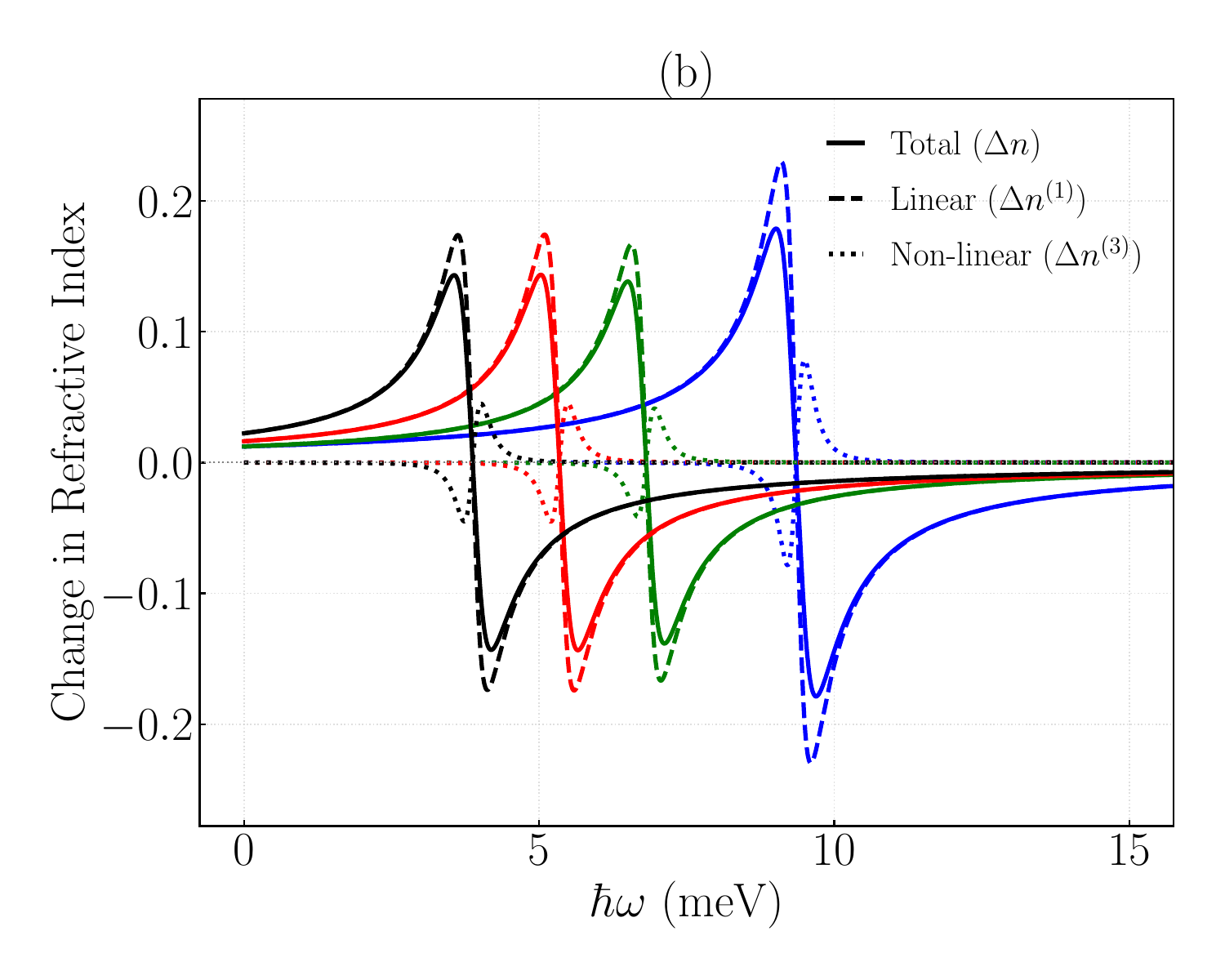}
    \caption{\footnotesize
    (Color online) Topological tuning of the low-energy $\Delta m=-1$ transition ($|+1\rangle \rightarrow |0\rangle$). (a) Absorption coefficient $\alpha(\omega)$ and (b) refractive index change $\Delta n(\omega)/n_{r}$ for different screw dislocation parameters $\beta$. Solid, dashed, and dotted lines indicate the total, linear, and third-order contributions, respectively. Fixed parameters: $I_{0}=2.2\times10^{6}$ W/m$^{2}$ and $\tau=1.0\times10^{7}$ m$^{-1}$.}
    \label{fig:beta_channel_pos}
\end{figure*}

In contrast, Fig.~~\ref{fig:beta_channel_pos} analyzes the low-energy absorption channel ($|+1\rangle \to |0\rangle$). Here, the screw dislocation produces a distinct impact: a systematic blueshift of the resonance peaks is observed as $\beta$ increases. This behavior indicates that the topological defect significantly raises the transition energy $\Delta E$ by altering the effective centrifugal term for this specific initial angular momentum state.

Thus, for the $|+1\rangle \to |0\rangle$ channel, $\beta$ acts primarily as a spectral controller. This fundamental difference between the two channels identifies the screw dislocation as a dual-purpose geometric tool: it provides spectral tuning for the low-energy transition while functioning as a robust amplitude modulator for the high-energy transition. This state-dependent response confirms the existence of geometrically induced, state-selective amplification in the system, in which the topological defect $\beta$ breaks the dynamical symmetry of the absorption pathways.

\subsection{Geometric Confinement and Nonlinear Switching (\texorpdfstring{$\tau$}{tau})}
\label{sec:tau_effects}

\begin{figure*}[tbhp]
    \centering
    \includegraphics[width=0.48\linewidth]{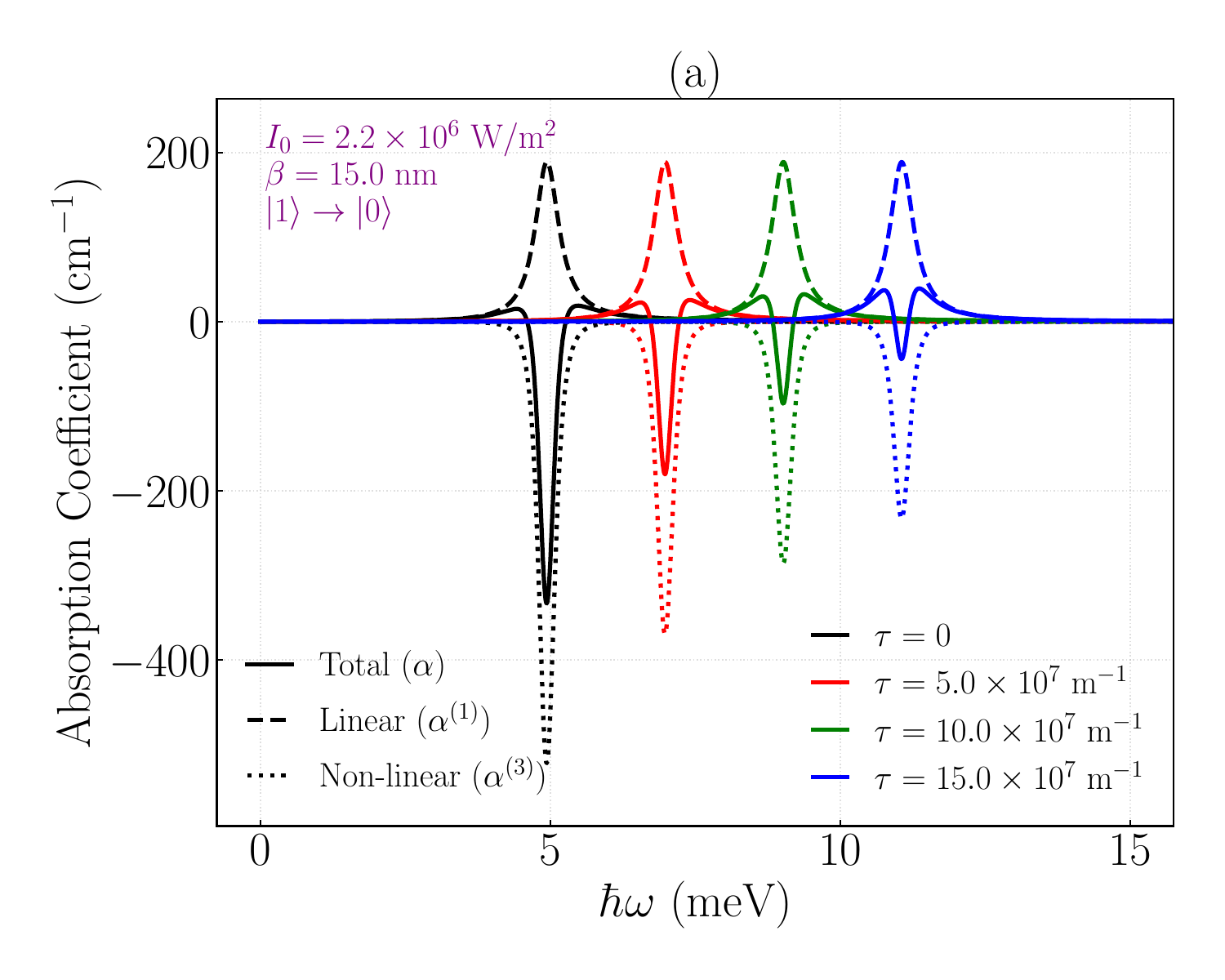}
    \includegraphics[width=0.48\linewidth]{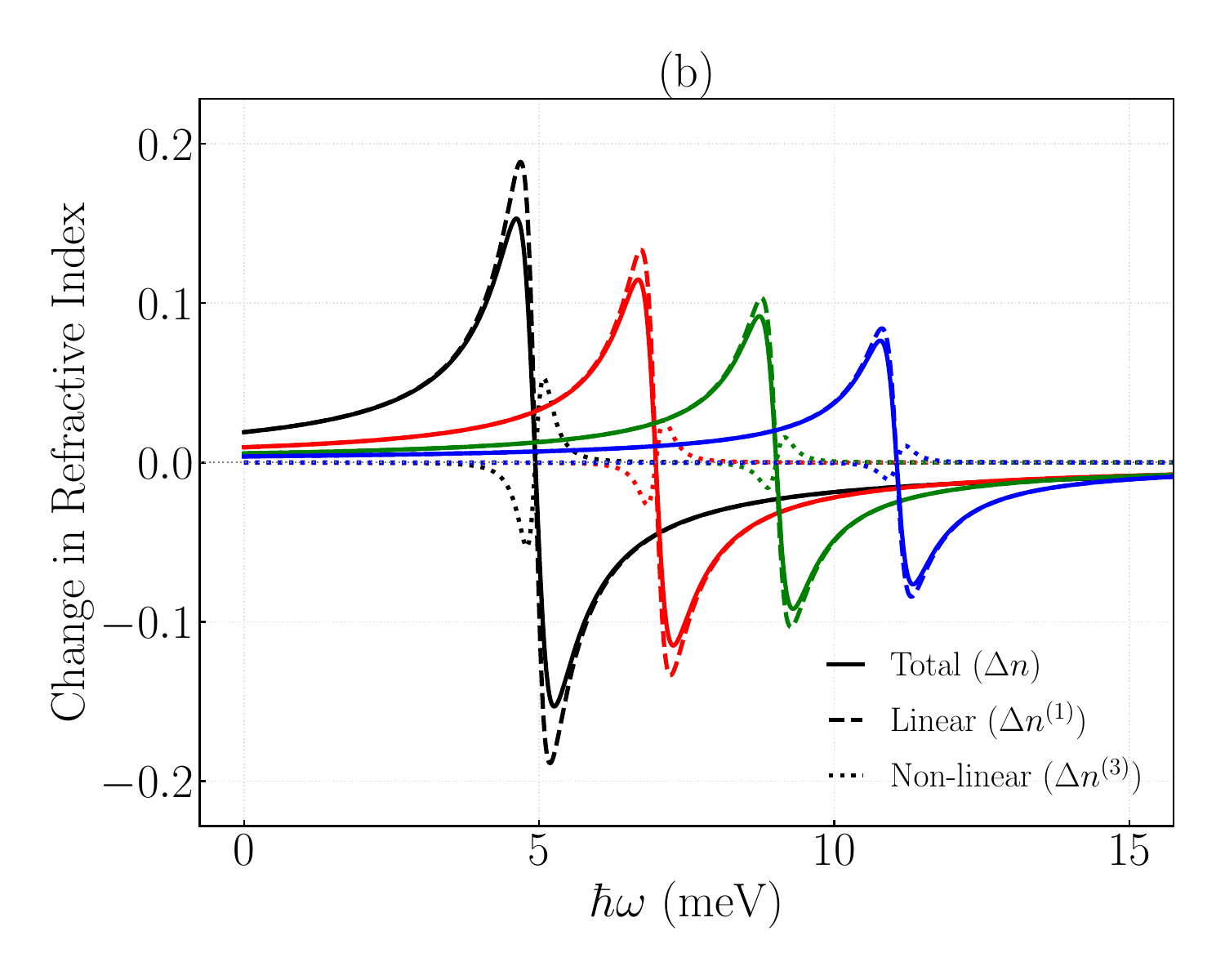}
    \caption{\footnotesize
    (Color online) Impact of geometric torsion on the optical response for the low-energy $\Delta m=-1$ transition ($|+1\rangle \rightarrow |0\rangle$). (a) Total absorption coefficient $\alpha(\omega)$ and (b) refractive index change $\Delta n(\omega)/n_{r}$ as functions of photon energy for varying torsion parameters $\tau$. Fixed parameters: $\beta=15.0$ nm and $I_{0}=2.2\times10^{6}$ W/m$^{2}$. }
    \label{fig:optical_tau_Io22_tau_m1}
\end{figure*}

Figure \ref{fig:optical_tau_Io22_tau_m1} presents the optical response for the low-energy transition channel ($|+1\rangle \to |0\rangle$) under a constant screw dislocation of $15$ nm. Panel (a) displays the linear, nonlinear, and total absorption coefficients, with the latter demonstrating the formation of an optical gain regime (negative values). Notably, as the torsion density $\tau$ increases, the resonance peaks undergo a pronounced blueshift, accompanied by a severe suppression of the gain magnitude. Physically, this occurs because the torsion enhances the effective harmonic confinement, which simultaneously increases the energy-level spacing and squeezes the radial wavefunctions toward the origin, thereby reducing the dipole matrix element. Consequently, panel (b) reflects this behavior in the refractive index change $\Delta n/n_r$: the typical anomalous dispersion curves track the spectral shift to higher energies, crossing the zero axis precisely at the new resonance frequencies, while their peak-to-peak amplitudes are strongly suppressed by the increasing torsion.

\begin{figure*}[tbhp]
    \centering
    \includegraphics[width=0.48\linewidth]{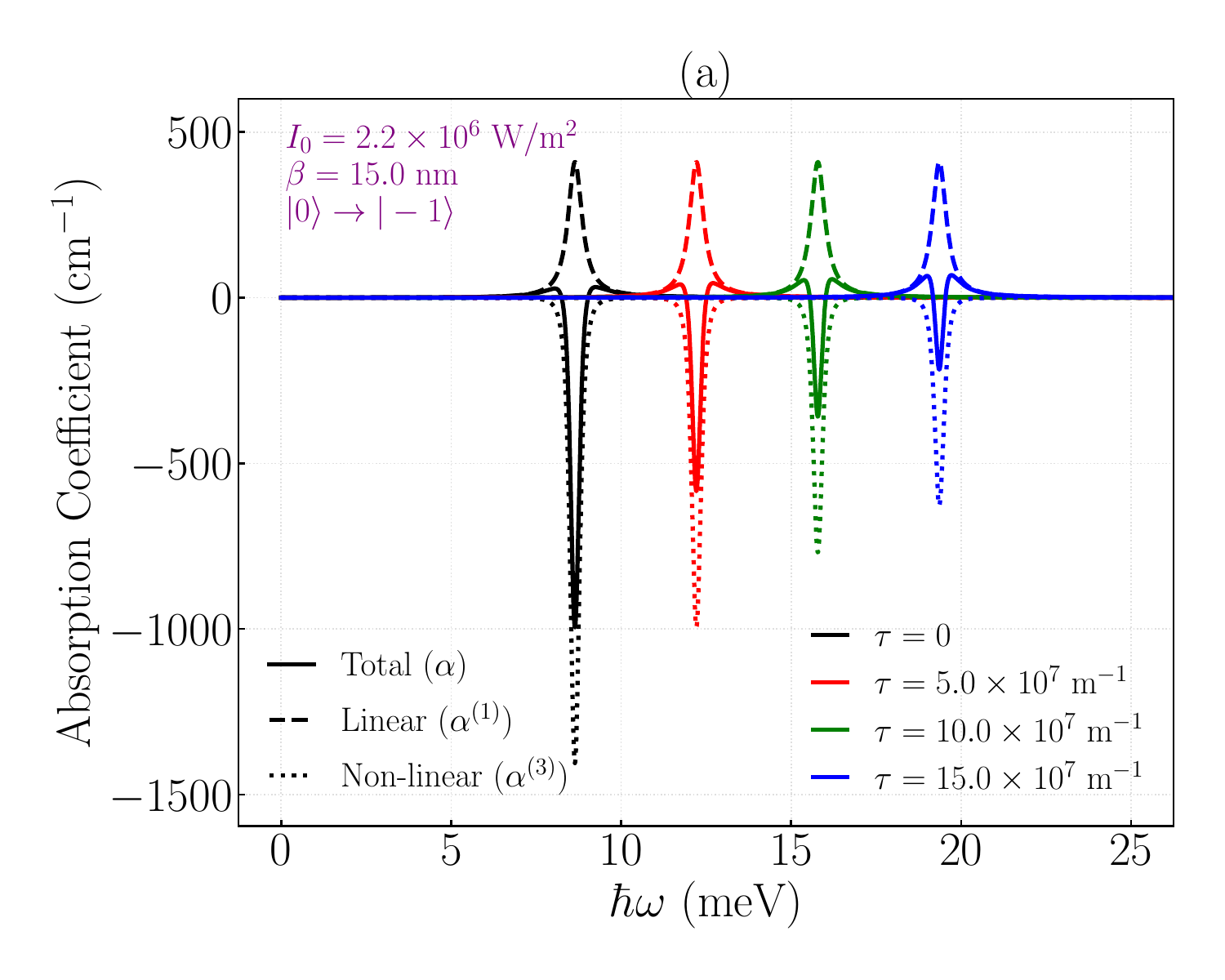}
    \includegraphics[width=0.48\linewidth]{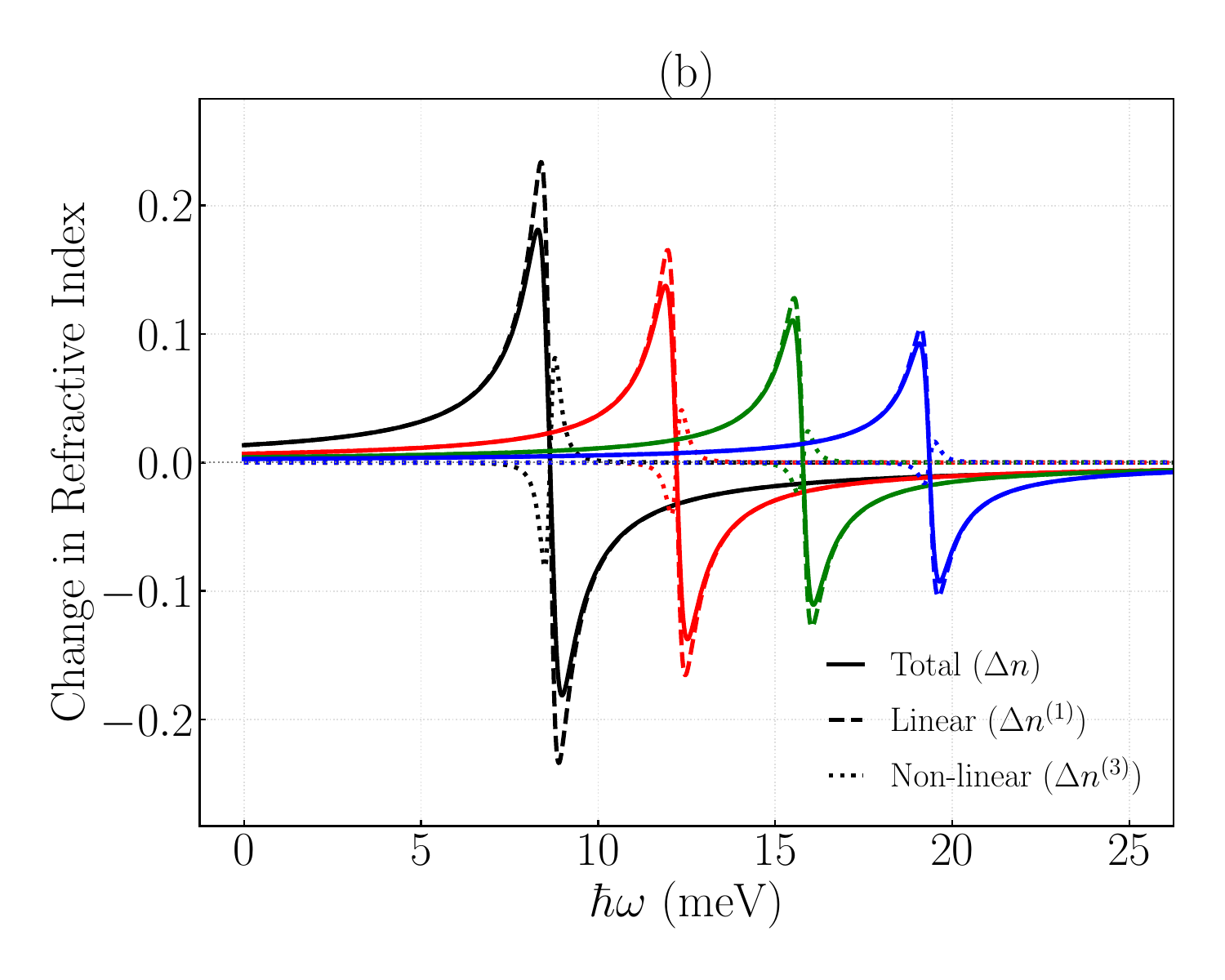}
    \caption{\footnotesize
    (Color online) Impact of geometric torsion on the optical response for the high-energy $\Delta m=-1$ transition ($|0\rangle \rightarrow |-1\rangle$). (a) Total absorption coefficient $\alpha(\omega)$ and (b) refractive index change $\Delta n(\omega)/n_{r}$ as functions of photon energy for varying torsion parameters $\tau$. Fixed parameters: $\beta=15.0$ nm and $I_{0}=2.2\times10^{6}$ W/m$^{2}$.}
    \label{fig:optical_tau_Io22_tau_m-1}
\end{figure*}

Figure \ref{fig:optical_tau_Io22_tau_m-1} illustrates the optical response for the high-energy channel $(|0\rangle \rightarrow|-1\rangle)$ in the presence of the topological defect ( $\beta=15 \mathrm{~nm}$ ). It is observed that both absorption and refractive index change occur at resonance points with considerably higher photon energies. Furthermore, the absorption coefficient exhibits significantly more intense optical gain peaks. However, the qualitative effect of torsion persists: as $\tau$ increases, the magnitude of this gain is progressively suppressed, reflecting the reduction in the dipole matrix element due to strong geometric confinement.

\subsection{Longitudinal quantum wire length $\left(L_z\right)$}
\begin{figure*}[tbhp]
    \centering
    \includegraphics[width=0.48\linewidth]{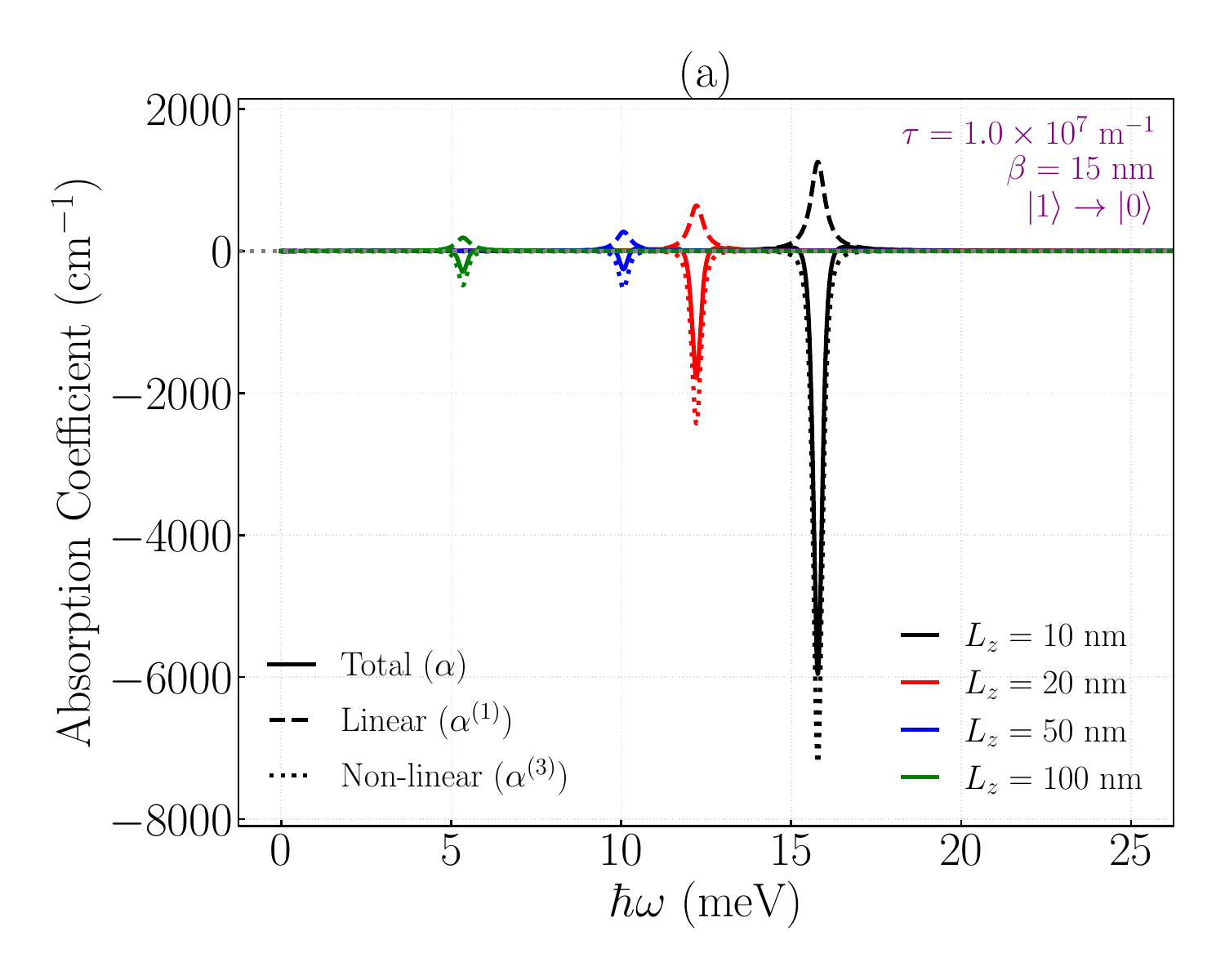}
    \includegraphics[width=0.48\linewidth]{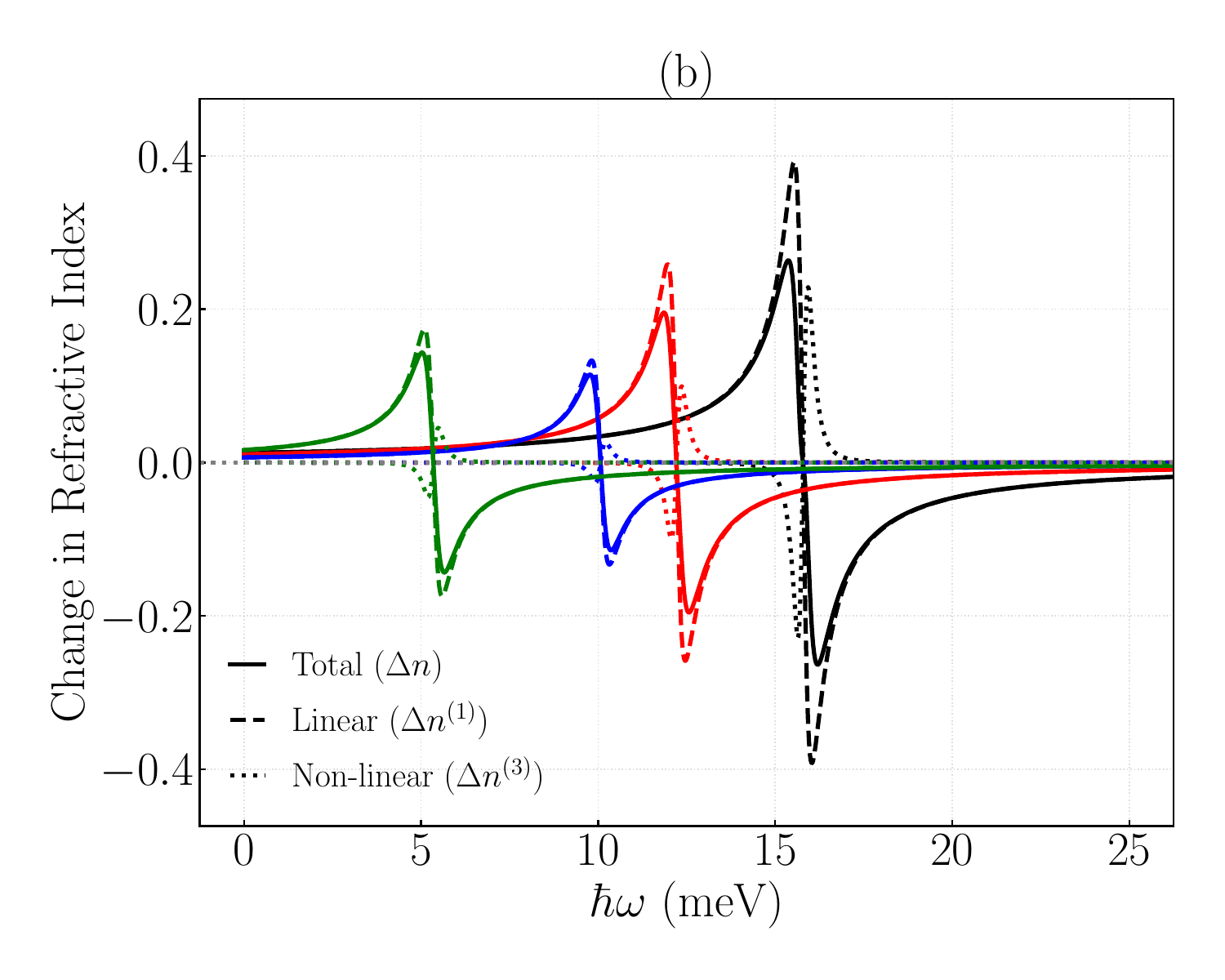}
    \caption{\footnotesize
    (Color online) Influence of the longitudinal quantum wire length $L_{z}$ on the optical response for the low-energy $\Delta m=-1$ transition ($|+1\rangle \rightarrow |0\rangle$). (a) Total absorption coefficient $\alpha(\omega)$ and (b) refractive index change $\Delta n(\omega)/n_{r}$ as functions of photon energy. Fixed parameters: $\tau=1.0\times10^{7}$ m$^{-1}$, $\beta=15.0$ nm, and $I_{0}=2.2\times10^{6}$ W/m$^{2}$.}
    \label{fig:optical_tau_Io22_tau_m1_b0}
\end{figure*}

Continuing our analysis, we investigate how the length of the quantum wire affects the system's optical properties. Fig.~\ref{fig:optical_tau_Io22_tau_m1_b0} illustrates the dependence of the optical response on the longitudinal length ($L_z$) for the low-energy transition channel ($|+1\rangle \to |0\rangle$). To isolate this effect, we kept the torsion density and screw dislocation fixed at $\tau = 1.0 \times 10^7$ m$^{-1}$ and $\beta = 15.0$ nm, under an incident intensity of $I_0 = 2.2 \times 10^6$ W/m$^2$.

Contrary to the effect observed when increasing the torsion, the elongation of $L_z$ (from 10 nm to 100 nm) induces a clear redshift in the resonance peaks. Simultaneously, a significant amplification in the magnitude of the optical gain is observed, with the negative absorption peaks in Fig.~\ref{fig:optical_tau_Io22_tau_m1_b0}(a) becoming significantly deeper for longer wires.

Physically, this behavior is governed by the quantization of the longitudinal momentum along the wire axis, given by $k_z = \pi/L_z$. As the wire elongates, the wave vector $k_z$ decreases, which directly weakens the contribution of the geometric coupling term ($k_z \tau$) to the effective harmonic confinement frequency. The relaxation of this radial confinement has two immediate consequences: first, it reduces the energy spacing between the transition levels, accounting for the spectral shift towards lower energies; second, it allows the electronic wavefunctions to expand spatially, moving further from the origin. This decompression favors an increase in the dipole matrix element, thereby enhancing transition efficiency and amplifying the nonlinear response. Fig.~\ref{fig:optical_tau_Io22_tau_m1_b0}(b) corroborates this dynamic, showing that the anomalous dispersion associated with the refractive index change also migrates to lower energies, with its peak-to-peak amplitude being drastically amplified in longer wires.

\begin{figure*}[tbhp]
    \centering
    \includegraphics[width=0.48\linewidth]{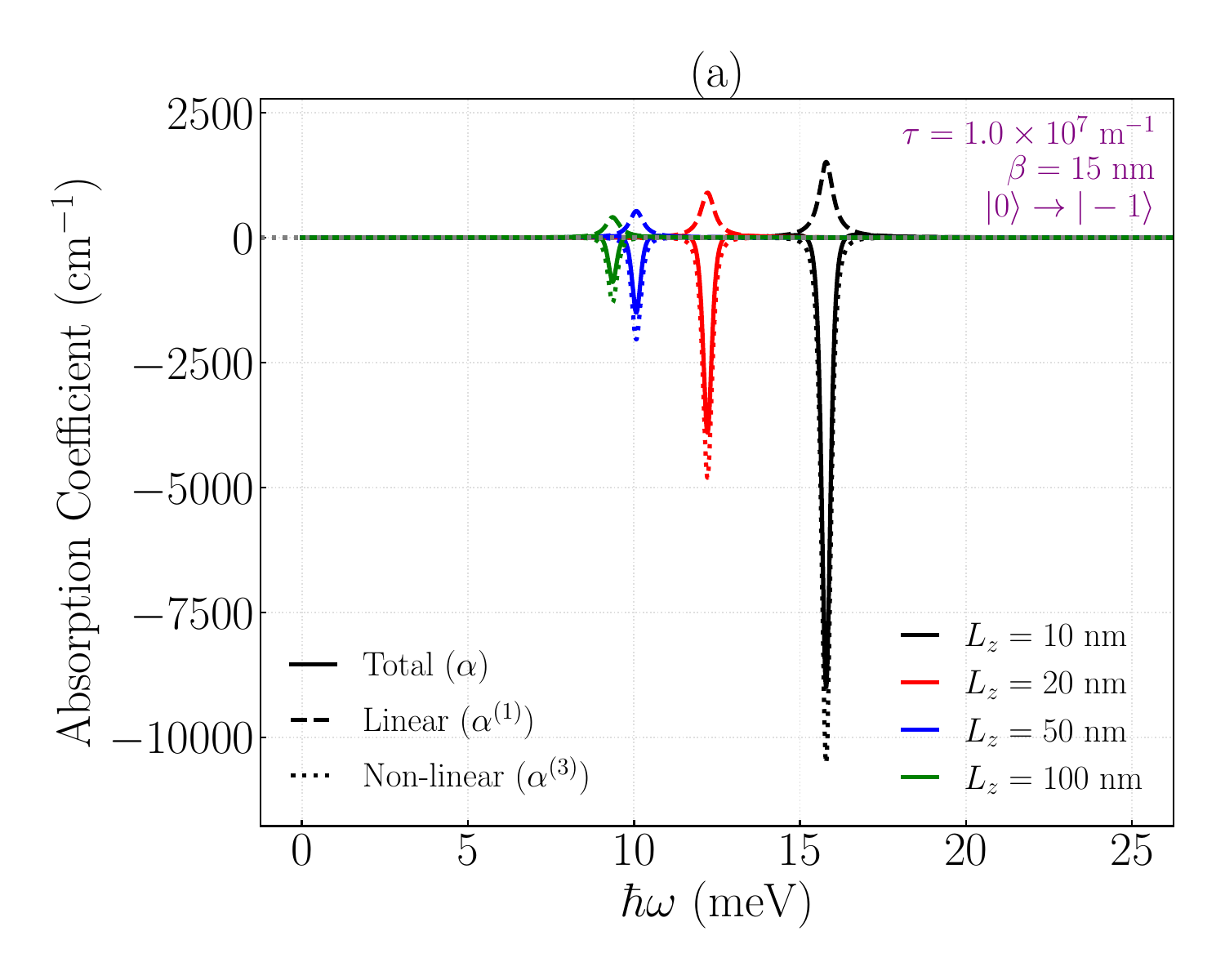}
    \includegraphics[width=0.48\linewidth]{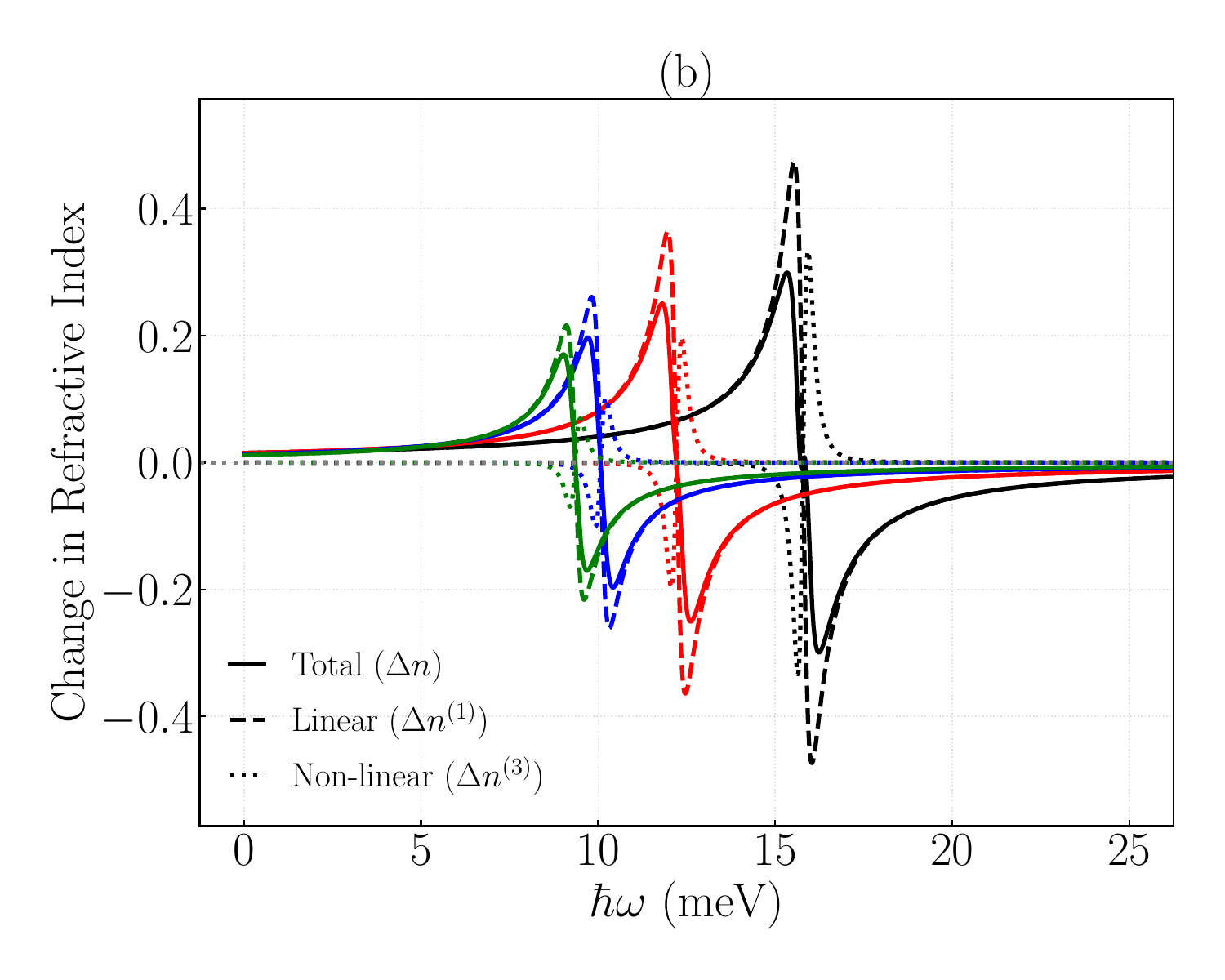}
    \caption{\footnotesize
    (Color online) Influence of the longitudinal quantum wire length $L_{z}$ on the optical response for the high-energy $\Delta m=-1$ transition ($|0\rangle \rightarrow |-1\rangle$). (a) Total absorption coefficient $\alpha(\omega)$ and (b) refractive index change $\Delta n(\omega)/n_{r}$ as functions of photon energy. Fixed parameters: $\tau=1.0\times10^{7}$ m$^{-1}$, $\beta=15.0$ nm, and $I_{0}=2.2\times10^{6}$ W/m$^{2}$.}
    \label{fig:optical_tau_Io22_tau_m-1_b10}
\end{figure*}

To evaluate the effect of the wire length on the second transition channel, Fig.~\ref{fig:optical_tau_Io22_tau_m-1_b10} presents the evolution of the optical response for the high-energy channel ($|0\rangle \to |-1\rangle$) as a function of photon energy for different longitudinal lengths $L_z$. The geometric constraints and incident intensity are kept the same ($\tau =$ 1.0 $\times$ 10$^7$ m$^{-1}$, $\beta =$ 15.0 nm, and $I_0 =$ 2.2 $\times$ 10$^6$ W/m$^2$).

The underlying physical mechanism remains the same, given that the elongation of the wire reduces the wave vector $k_z$ and weakens the torsion-induced radial confinement. However, the results reveal a striking quantitative contrast between the two transition channels. As in the previous case, increasing $L_z$ induces a clear redshift in the resonances. Nevertheless, due to the constructive interaction between the angular momentum of this state and the topological defect $\beta$, the transition not only occurs in a much higher-energy spectral window, but it also results in a significantly more pronounced enhancement of the dipole matrix element due to the relaxation of the confinement.

The underlying physical mechanism remains the same, given that the elongation of the wire reduces the wave vector $k_z$ and weakens the torsion-induced radial confinement. However, the results reveal a striking quantitative contrast between the two transition channels. As in the previous case, increasing $L_z$ induces a clear redshift in the resonances. It is noted, however, that this spectral shift undergoes a saturation for larger values of $L_z$ (blue and green curves tending to cluster together). This asymptotic behavior occurs because $k_z \propto 1/L_z$; for very long wires, the geometric coupling term $k_z \tau$ tends to zero, and the resonance position converges to a limit dictated predominantly by the magnetic field and the topological defect. Furthermore, due to the constructive interaction between the angular momentum of this state and $\beta$, the transition not only occurs in a much higher-energy spectral window, but it also results in a significantly more pronounced enhancement of the dipole matrix element due to the relaxation of the confinement.

The most striking impact of this weakened confinement is reflected in the transition efficiency. As evidenced in Fig.~\ref{fig:optical_tau_Io22_tau_m-1_b10}(a), the increase in the dipole matrix element for longer wires strongly intensifies the magnitude of the nonlinear optical gain, with the peaks reaching absolute values orders of magnitude greater than those observed in the low-energy channel. Fig.~\ref{fig:optical_tau_Io22_tau_m-1_b10}(b) corroborates this dynamic: the anomalous dispersion of the refractive index maintains the same profile as the previous case, but its peak-to-peak amplitudes become considerably larger. This behavior establishes the wire length $L_z$ not merely as a spectral tuner but as a fundamental mechanism for maximizing state-selective gain.

\subsection{Photoionization Cross Section}
\begin{figure}
    \centering
    \includegraphics[width=1\linewidth]{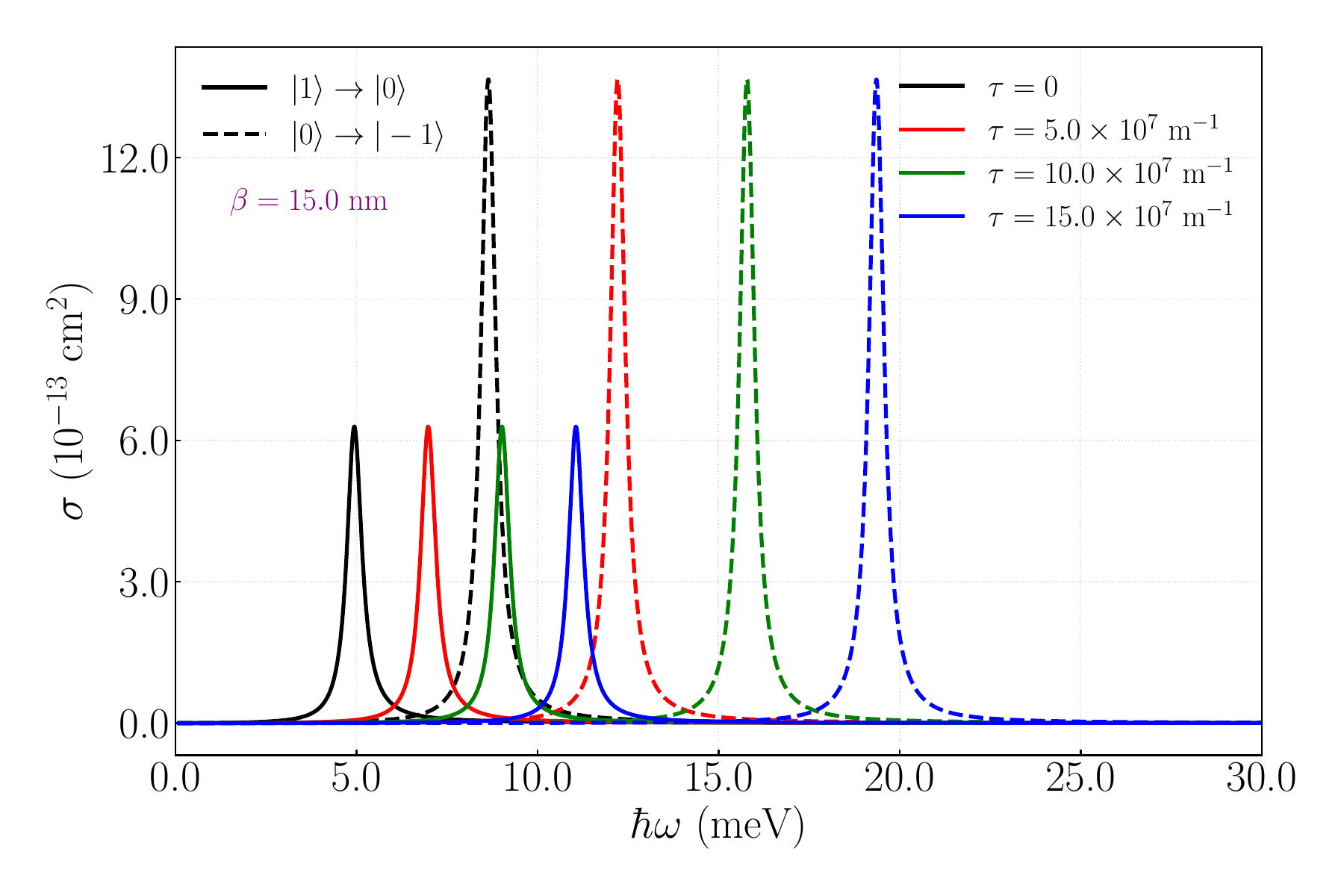}
    \caption{\footnotesize
    (Color online) Photoionization cross section $\sigma$ versus photon energy $\hbar\omega$ for the dipole-allowed $\Delta m=-1$ transitions. Solid and dashed lines correspond to the $|+1\rangle \rightarrow |0\rangle$ and $|0\rangle \rightarrow |-1\rangle$ channels, respectively, evaluated for different torsion densities $\tau$. The screw dislocation is fixed at $\beta=15.0$ nm.}
    \label{fig:photoionization_tau}
\end{figure}

The photoionization cross-section equation (Eq.~\ref{eq:sigma_working_main}) reflects the same trend observed in bound-bound absorption. Figure~\ref{fig:photoionization_tau} corroborates the state-dependent nature of the optical response by investigating the photoionization cross section (PCS) for the $\Delta m=-1$ channels. By fixing the topological defect at $\beta = 15.0$~nm, we identify the optical fingerprint of the torsion-induced confinement through two distinct physical signatures. First, increasing the torsion density pushes the resonance toward higher photon energies (a clear blueshift) for both transition channels. This shift is accompanied by a simultaneous suppression of the PCS peak amplitude, reflecting the compression of the radial wave functions toward the origin and the consequent reduction in the dipole matrix overlap. Second, the spectral asymmetry dictated by the geometric breaking of dynamical symmetry becomes evident: the high-energy channel $(|0\rangle\rightarrow|-1\rangle)$ emerges at substantially higher energies and with greater intensity due to the constructive interference between the topological defect and the effective angular momentum. This contrasts with the partial cancellation, and consequent emergence at lower energies, observed in the low-energy channel $(|+1\rangle\rightarrow|0\rangle)$

\begin{figure}
    \centering
    \includegraphics[width=1\linewidth]{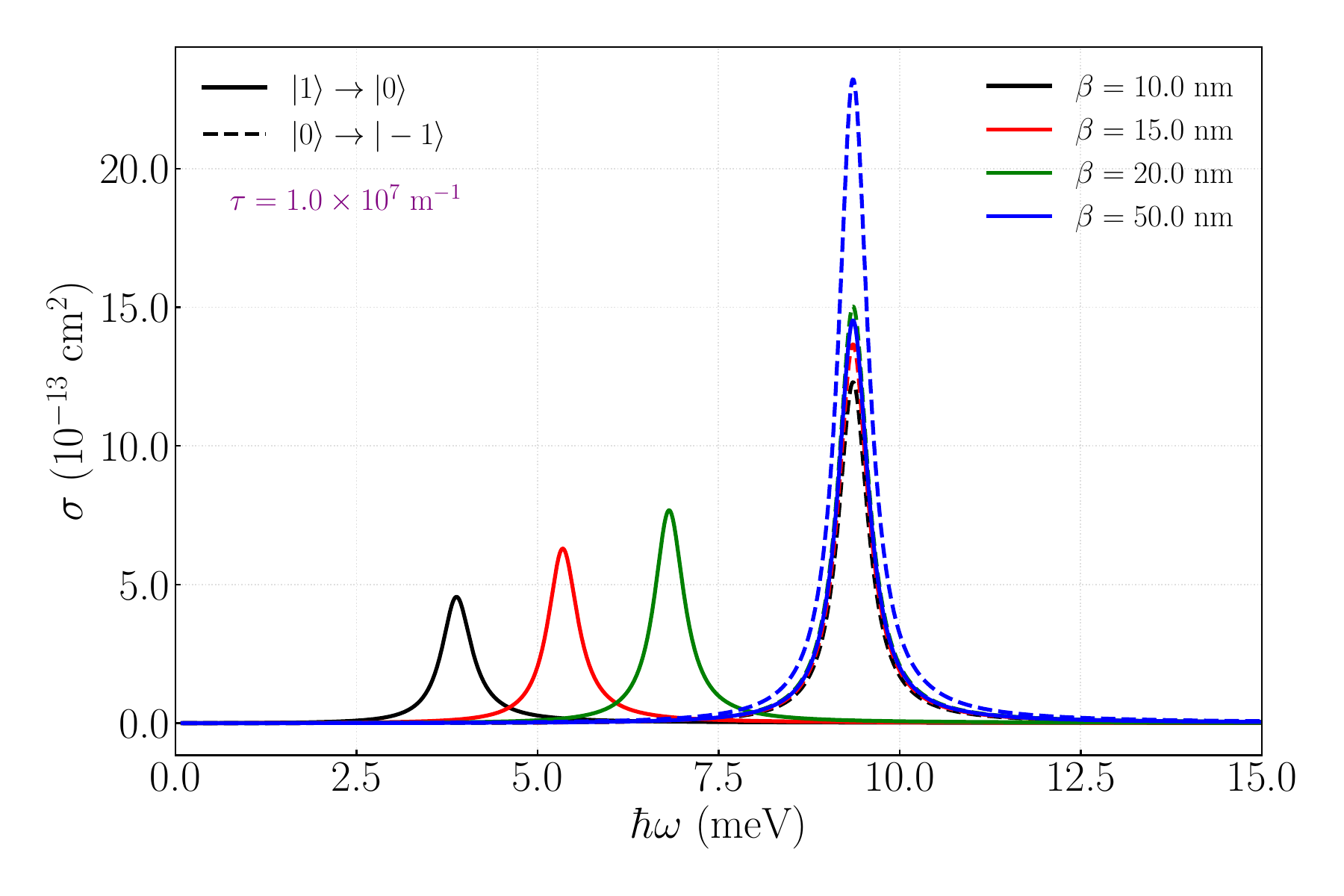}
    \caption{\footnotesize
    (Color online) Photoionization cross section $\sigma$ versus photon energy $\hbar\omega$ for the dipole-allowed $\Delta m=-1$ transitions. Solid and dashed lines correspond to the $|+1\rangle \rightarrow |0\rangle$ and $|0\rangle \rightarrow |-1\rangle$ channels, respectively, evaluated for different screw dislocation parameters $\beta$. The torsion density is fixed at $\tau=1.0\times10^{7}$ m$^{-1}$.}
    \label{fig:photoionization_beta}
\end{figure}

On the other hand, investigating the screw dislocation parameter $\beta$ in the photoionization cross section reveals a highly selective spectral control mechanism, as illustrated in Fig.~\ref{fig:photoionization_beta}. Keeping the torsion density fixed, it is observed that increasing $\beta$ induces a clear blueshift exclusively in the low-energy transition channel ($|+1\rangle \rightarrow |0\rangle$). In contrast, the resonance position for the high-energy channel ($|0\rangle \rightarrow |-1\rangle$) remains unchanged. Physically, this spectral robustness arises because the topological shift term in the effective angular momentum cancels out in the energy difference $\Delta E$ between these two specific states. Furthermore, alongside this asymmetric spectral tuning, the results demonstrate that increasing $\beta$ amplifies the peak magnitudes of the PCS in both channels. This amplitude enhancement reflects a reconfiguration of the centrifugal barrier that favors the spatial overlap of the wave functions, indicating that the topological defect globally increases the optical transition probability.

\begin{figure}
    \centering
    \includegraphics[width=1\linewidth]{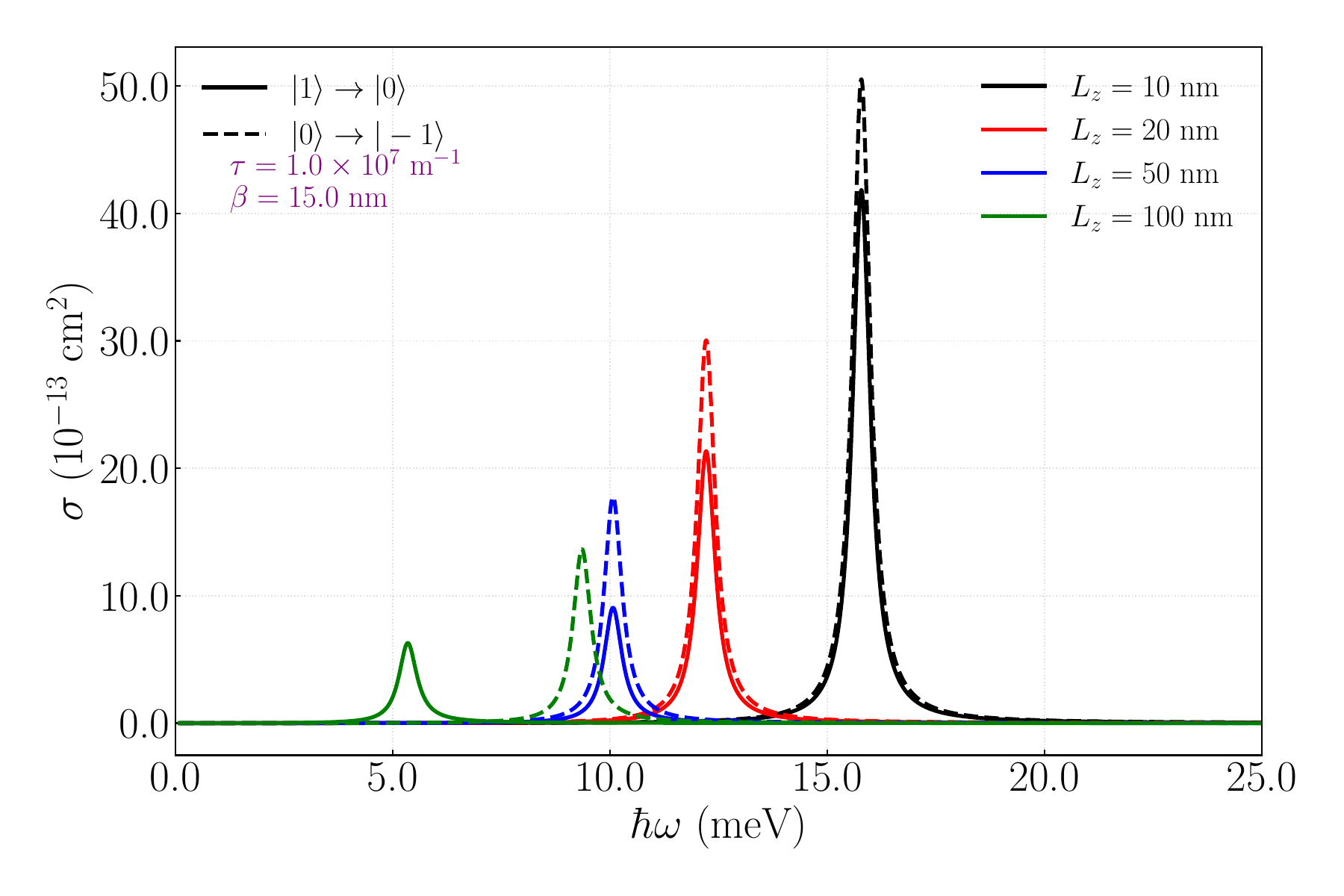}
    \caption{\footnotesize
    (Color online) Photoionization cross section $\sigma$ versus photon energy $\hbar\omega$ for the dipole-allowed $\Delta m=-1$ transitions, evaluating the effect of the quantum wire length $L_{z}$. Solid and dashed lines correspond to the $|+1\rangle \rightarrow |0\rangle$ and $|0\rangle \rightarrow |-1\rangle$ channels, respectively. Fixed geometric parameters: $\tau=1.0\times10^{7}$ m$^{-1}$ and $\beta=15.0$ nm.}
\label{fig:photoionization_lz}
\end{figure}

To conclude the analysis of the photoionization cross section, Fig.~\ref{fig:photoionization_lz} investigates the influence of the quantum wire's longitudinal length ($L_z$), keeping the geometric parameters fixed at $\tau = 1.0 \times 10^7$~m$^{-1}$ and $\beta = 15.0$~nm. Since the torsion coupling in the effective potential scales with the longitudinal momentum ($k_z \tau$, where $k_z = \pi/L_z$), varying $L_z$ directly modulates the strength of this confinement. The plot demonstrates that elongating the wire (from 10~nm to 100~nm) weakens the effective radial confinement, inducing a pronounced redshift in the resonance peaks of both transition channels. Specifically, the high-energy channel ($|0\rangle \rightarrow |-1\rangle$) shifts back from approximately 16~meV to nearly 9~meV. Simultaneously, a suppression of the maximum PCS amplitude is observed as the wire becomes longer. Although the relaxation of the radial confinement allows for the spatial expansion of the wave functions (thereby increasing the dipole matrix element), the linear dependence of the cross section on the incident photon energy ($\hbar\omega$) dominates the optical response in this regime. This ultimately attenuates the global photoionization probability for longer wires. The strong spectral and amplitude asymmetry between the two $\Delta m=-1$ channels is preserved over the entire range, thereby confirming the state-selective optical response governed by the system's geometry.

\subsection{Oscillator strength and its torsion/topology dependence}
\label{sec:osc_strength}

The (dimensionless) oscillator strength quantifies the strength of an electric-dipole transition between two eigenstates. In the length gauge, it is defined as
\begin{equation}
f_{fi}
=
\frac{2 m^{*}}{\hbar^{2}}\,
\big(E_f-E_i\big)\,
\big|\langle \psi_f |\, \hat{\boldsymbol r}\cdot\hat{\boldsymbol e}_r\,| \psi_i \rangle\big|^{2},
\label{eq:def_f_fi}
\end{equation}
where $\hat{\boldsymbol e}_r$ is the in-plane polarization vector and $E_f-E_i\equiv\Delta E$ is the transition energy obtained from the analytic spectrum in Eq.~\eqref{eq:energy_eigenvalues}.

For our cylindrically symmetric states
$\psi_{n,m,k_z}(\rho,\varphi,z)= \frac{1}{\sqrt{2\pi}} R_{n,m}(\rho)\,e^{i m\varphi}\,e^{i k_z z}$,
only $\Delta m=\pm 1$ dipole transitions are allowed. Integrating over the angular coordinates, the dipole matrix element reduces to a purely radial overlap,
\begin{align}
M_{fi}
&=
\int_{0}^{\infty}
R_{n',m\pm1}(\rho)\,\rho^{2}\,R_{n,m}(\rho)\, d\rho,
\label{eq:Mfi_radial_main}
\end{align}
where $R_{n,m}(\rho)$ are the normalized radial wavefunctions from Eq.~\eqref{eq:wavefunction_explicit}.

It is convenient to express this overlap in terms of the dimensionless variable
$\xi = |\Omega| \rho^2$, where
$|\Omega| = |eB/2\hbar + k_z \tau|$
is the effective confinement frequency introduced in Sec.~\ref{sec:helical_metric}.
After inserting Eq.~\eqref{eq:wavefunction_explicit} and changing variables $\rho\mapsto\xi$, the integral in Eq.~\eqref{eq:Mfi_radial_main} becomes
\begin{align}
M_{fi}
&=
\frac{1}{2 |\Omega|^{\frac{|j_f|+|j_i|+3}{2}}}\,
\mathcal{N}_{n',j_f}\,\mathcal{N}_{n,j_i} \notag \\& \times
\int_{0}^{\infty}
\xi^{\frac{|j_i|+|j_f|+1}{2}}
e^{-\xi}\,
L_{n'}^{(|j_f|)}(\xi)\,
L_{n}^{(|j_i|)}(\xi)\,
d\xi,
\label{eq:Mfi_dimless_main}
\end{align}
where the effective angular momenta are $j_i = m - l - k_z\beta$ and $j_f = m\pm 1 - l - k_z\beta$, $L_{n}^{(|j|)}$ are associated Laguerre polynomials, and the normalization constant is
\begin{equation}
\mathcal{N}_{n,j}
=
\sqrt{\frac{2|\Omega|^{|j|+1}\,n!}{\Gamma(n+|j|+1)}}.
\end{equation}

The remaining Laguerre overlap integral is standard. Its closed form can be tabulated analytically or evaluated numerically. Inserting $M_{fi}$ from Eq.~\eqref{eq:Mfi_dimless_main} into Eq.~\eqref{eq:def_f_fi} gives $f_{fi}$ for any allowed $(n,m)\to(n',m\pm1)$ channel.

Two physical ingredients determine $f_{fi}$:
1. The transition energy $\Delta E = E_f - E_i$, which increases with the effective confinement frequency $\omega_{eff}$ [see Eq.~\eqref{eq:energy_eigenvalues}]. Here $\omega_{eff} = \frac{eB}{m^*} + \frac{2\hbar k_z \tau}{m^*}$ encompasses both the purely magnetic cyclotron contribution (from the perpendicular field $B$) and the geometric contribution induced by the torsion density ($\tau$) coupled to the longitudinal motion ($k_z$).

2. The spatial overlap $|M_{fi}|$, which decreases as torsion and the magnetic field squeeze the electronic wavefunctions radially. Both $\tau$ and $B$ enter the confinement scale $|\Omega|=|eB/2\hbar + k_z\tau|$; increasing $|\Omega|$ tightens the radial probability density and tends to reduce $|M_{fi}|$.

The oscillator strength $f_{fi}\propto \Delta E\,|M_{fi}|^2$ is therefore the result of a tug-of-war: torsion and field push the levels apart (larger $\Delta E$) but at the same time reduce their spatial overlap (smaller $|M_{fi}|$).

\begin{figure}[htpb]
    \centering
    \includegraphics[width=\linewidth]{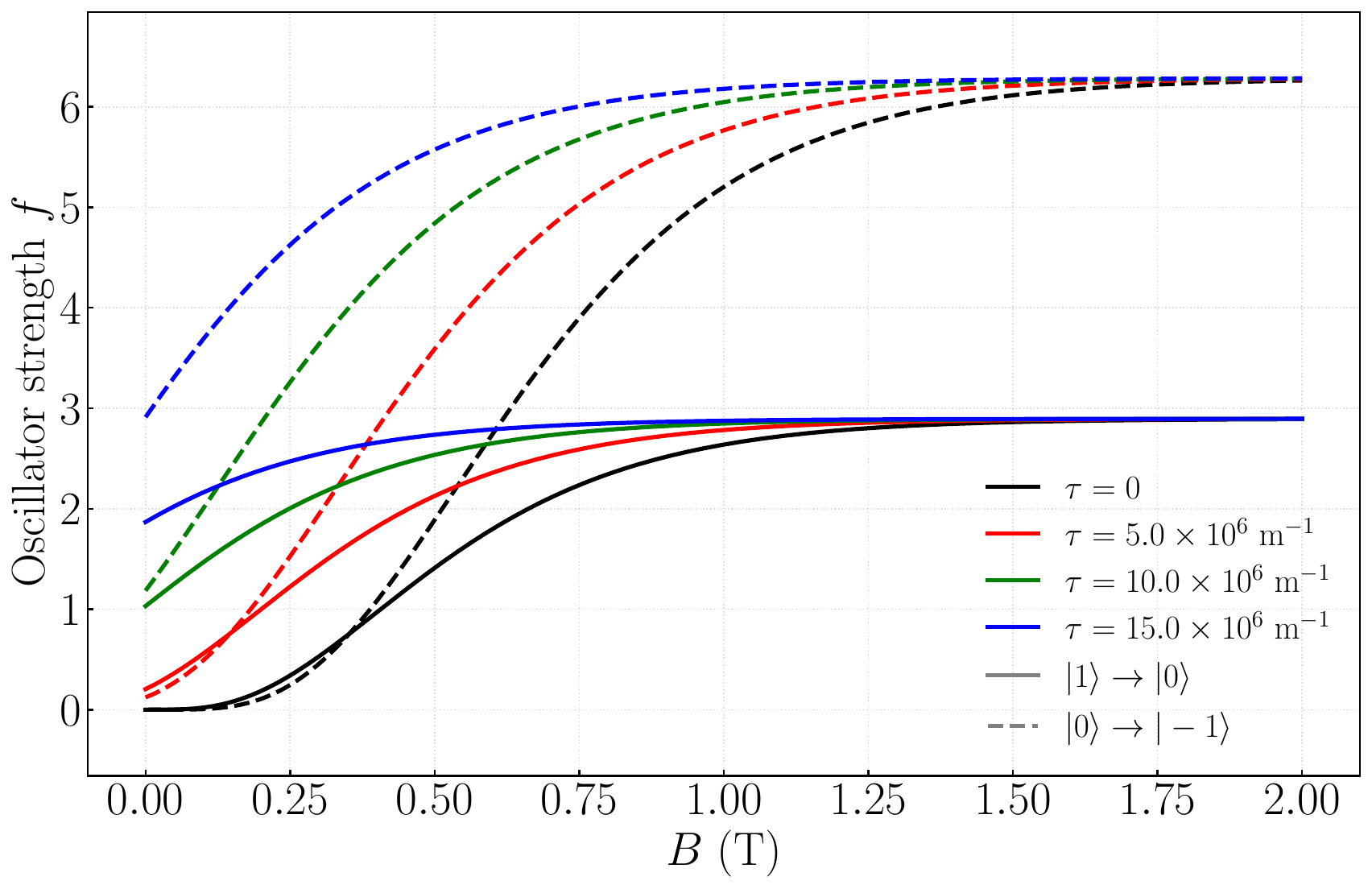}
    \caption{\footnotesize
    (Color online) Oscillator strength $f$ as a function of the perpendicular magnetic field $B$ for various torsion densities $\tau$. Solid and dashed lines correspond to the low-energy ($|1\rangle \rightarrow |0\rangle$) and high-energy ($|0\rangle \rightarrow |-1\rangle$) transition channels, respectively. Fixed system parameters: $\beta=15.0$ nm, $l=0.1$, and $L_{z}=100$ nm. }
    \label{fig:ftau}
\end{figure}

This interplay is explicitly depicted in Fig.~\ref{fig:ftau}, which shows the magnetic field dependence of the oscillator strength for different torsion densities ($\tau$). As the magnetic field $B$ increases from zero, the oscillator strength initially grows rapidly before saturating at a constant value for higher fields. This saturation occurs because, in the strong magnetic field limit ($eB/2\hbar \gg k_z\tau$), the effective confinement frequency $\Omega$ is dominated by the cyclotron contribution, causing both the energy spacing $\Delta E$ and the radial overlap $|M_{fi}|$ to stabilize. Furthermore, a pronounced state-selective asymmetry is observed: the high-energy channel $|0\rangle\rightarrow|-1\rangle$ (dashed lines) reaches a significantly higher saturation value than the low-energy channel $|1\rangle\rightarrow|0\rangle$ (solid lines). Regarding the effect of torsion, increasing $\tau$ systematically suppresses the overall oscillator strength in both channels. Moreover, higher torsion densities delay the onset of the saturation regime, requiring stronger magnetic fields to stabilize the transition efficiency. Ultimately, this indicates that the spatial squeezing of the wavefunctions (the reduction in $|M_{fi}|^2$) dominates over the geometrically induced blueshift ($\Delta E$ increase), effectively quenching the optical response.

\begin{figure}[htpb]
    \centering
    \includegraphics[width=\linewidth]{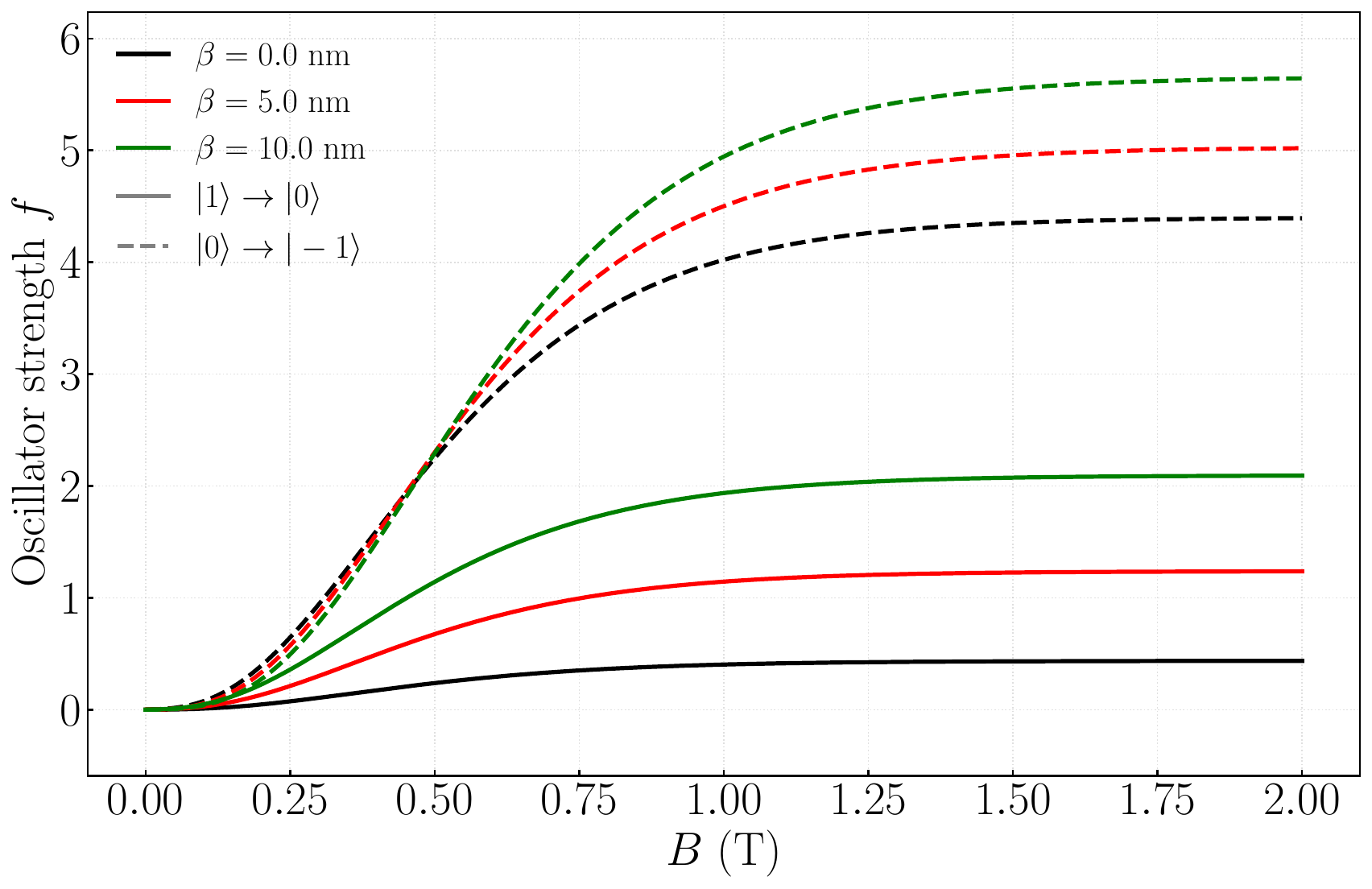}
    \caption{\footnotesize
    (Color online) Oscillator strength $f$ as a function of the perpendicular magnetic field $B$ for various screw dislocation parameters $\beta$. Solid and dashed lines correspond to the low-energy ($|1\rangle \rightarrow |0\rangle$) and high-energy ($|0\rangle \rightarrow |-1\rangle$) transition channels, respectively. Fixed geometric parameters: $\tau=1.0\times10^{6}$ m$^{-1}$, $l=0.1$, and $L_{z}=100$ nm.}
    \label{fig:fbeta}
\end{figure}
\begin{figure}[htpb]
    \centering
    \includegraphics[width=\linewidth]{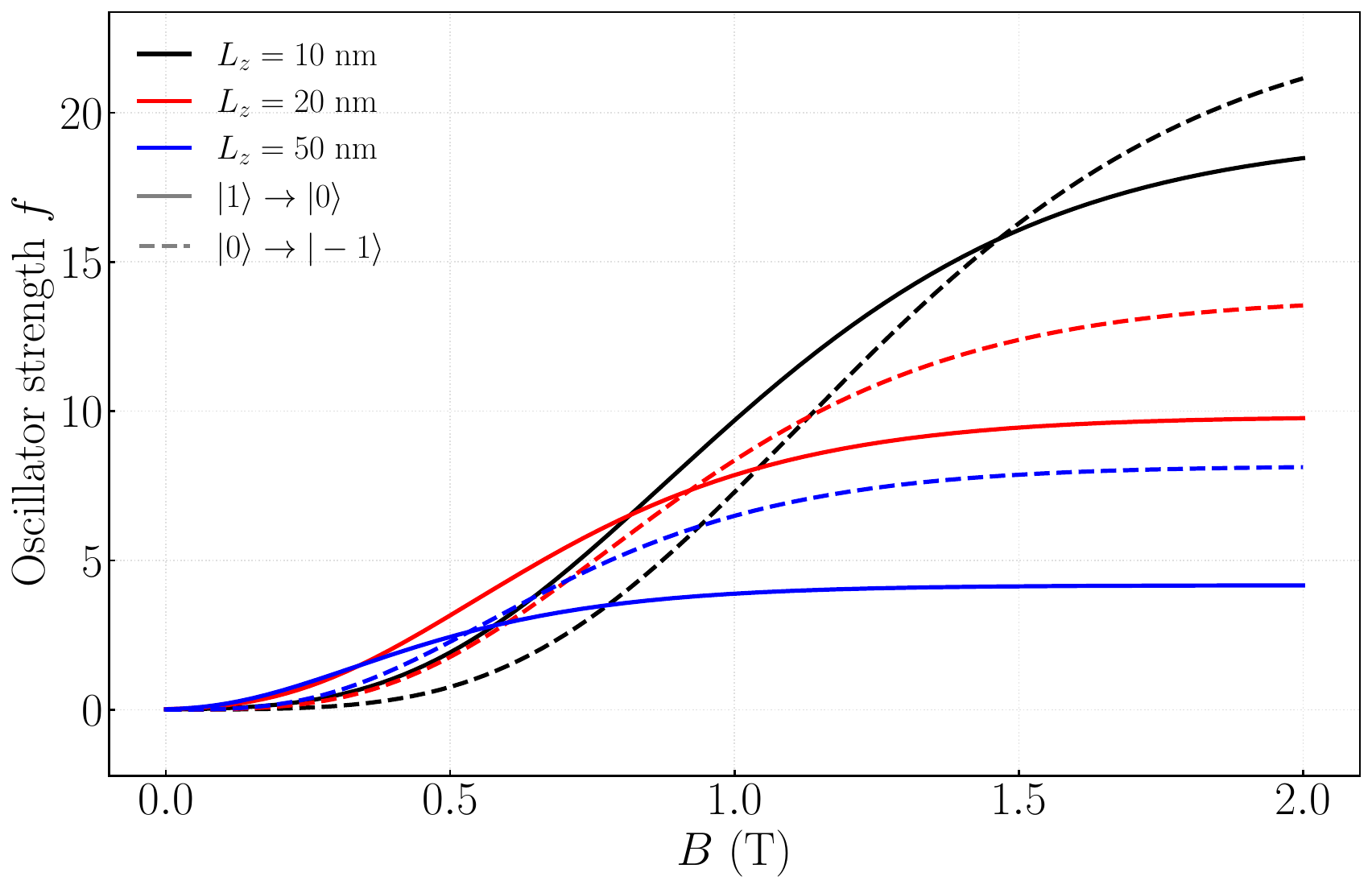}
   \caption{\footnotesize
   (Color online) Oscillator strength $f$ as a function of the perpendicular magnetic field $B$ for different longitudinal quantum wire lengths $L_{z}$. Solid and dashed lines correspond to the low-energy ($|1\rangle \rightarrow |0\rangle$) and high-energy ($|0\rangle \rightarrow |-1\rangle$) transition channels, respectively. Fixed geometric parameters: $\tau=1.0\times10^{6}$ m$^{-1}$, $\beta=15.0$ nm, and $l=0.1$.}
    \label{fig:fl}
\end{figure}

In contrast, the influence of the screw dislocation parameter ($\beta$) on the oscillator strength is shown in Fig.~\ref{fig:fbeta}. As observed, the oscillator strength increases with the magnetic field $B$ until it saturates at a constant value. Notably, the high-energy channel ($|0\rangle \rightarrow |-1\rangle$, dashed lines) saturates at a significantly higher value than the low-energy channel ($|1\rangle \rightarrow |0\rangle$, solid lines). The primary role of $\beta$ is to order these saturation levels: an increase in $\beta$ directly increases the saturation value of the oscillator strength.

Finally, Fig.~\ref{fig:fl} investigates the role of the quantum wire length ($L_z$). As the wire elongates (e.g., from 10 nm to 100 nm), the longitudinal wave vector $k_z = \pi/L_z$ decreases, which weakens the torsion-induced geometric confinement term $k_z \tau$. This relaxation allows the electronic wavefunctions to decompress and expand radially, thereby driving a substantial enhancement of the dipole matrix element that overwhelms the concurrent reduction in transition energy. Consequently, longer wires yield drastically amplified oscillator strengths. Furthermore, the plot reveals a distinct, field-dependent crossing behavior between the two channels. For any given wire length, the low-energy channel ($|1\rangle \rightarrow |0\rangle$, solid lines) exhibits a stronger optical response at low magnetic fields. However, as $B$ increases, a crossover occurs, and the high-energy channel ($|0\rangle \rightarrow |-1\rangle$, dashed lines) overtakes it, eventually saturating at a higher level. Notably, increasing $L_z$ not only amplifies the overall magnitude but also shifts this geometric crossing point to significantly lower magnetic fields, offering an additional parameter to tune state-selective absorption.

\section{Experimental relevance and spectroscopic fingerprints}
\label{sec:experimental_relevance}

The results obtained in this work are directly connected to experimentally accessible spectroscopic observables and therefore admit a natural interpretation in terms of measurable fingerprints of torsion-induced confinement and topological symmetry breaking. Rather than relying on indirect indicators, the present model predicts clear modifications in quantities that can be probed in standard mid-infrared and terahertz spectroscopies, including the absorption coefficient, nonlinear bleaching and gain onset, refractive-index dispersion, photoionization cross section, and oscillator strength \cite{SheikBahae1990OL,SheikBahae1991JOSAB,ChemlaShah2001Nature,Dinu2003APL,Autere2018AdvMater,Kippenberg2018Science,Jin2018SciRep_QD_PCS_Exp,Kuyucu2014SST_QD_PCS_Review}. In this sense, the relevance of the present system lies in the fact that each geometric ingredient leaves a distinct and experimentally traceable signature on the optical spectrum.

The first robust signature is the displacement of the resonance energies. Since the torsion density $\tau$ increases the effective radial confinement through its coupling to the longitudinal motion, it systematically drives the resonances toward higher photon energies. This blueshift appears consistently in the linear and nonlinear absorption spectra, in the dispersive zero crossing of the refractive-index change, and in the photoionization cross section. By contrast, increasing the quantum-wire length $L_z$ reduces the longitudinal wave vector and weakens the corresponding geometric confinement mechanism, producing the opposite trend: a redshift of the spectral features. This complementary behavior offers a particularly clear experimental strategy: by tracking the motion of the resonance peaks as a function of structural geometry, one may identify whether the dominant effect is enhanced torsional confinement or its relaxation through longitudinal delocalization.

A second measurable signature is the modification of the peak amplitudes. The torsion-induced squeezing of the radial wave functions reduces the dipole overlap between initial and final states and therefore suppresses the magnitude of the optical resonances. This attenuation is visible not only in the linear absorption peak, but also in the nonlinear contribution, the refractive-index modulation, the photoionization cross section, and the oscillator strength. In contrast, the screw dislocation parameter $\beta$ acts primarily through the effective angular momentum and the associated centrifugal barrier. Its role is therefore not to create confinement, but to redistribute the radial probability density and reshape the transition strength in a channel-dependent fashion. Experimentally, this distinction is important because it allows one to separate the compressive geometric action of torsion from the topological redistribution induced by the screw defect by analyzing line position and line intensity together.

The most distinctive fingerprint predicted here is the pronounced asymmetry between the two dipole-allowed $\Delta m=-1$ channels. Even though both transitions obey the same selection rule, the combined action of torsion, screw dislocation, magnetic field, and AB flux breaks the dynamical equivalence between opposite angular-momentum sectors. As a consequence, the channels $|+1\rangle \rightarrow |0\rangle$ and $|0\rangle \rightarrow |-1\rangle$ emerge at markedly different photon energies and with substantially different spectral weights. This asymmetry should be directly observable in channel-resolved absorption or photoexcitation measurements as a splitting between resonances that cannot be attributed to a simple symmetric magnetic confinement picture. In addition, under sufficiently intense optical driving, the same asymmetry extends to the nonlinear regime, where one channel may display much stronger bleaching or even negative absorption than the other. The observation of such channel-selective nonlinear behavior would provide a particularly strong indication of the geometric origin of the confinement mechanism.

The refractive response offers a further complementary diagnostic. Because the dispersive profile of $\Delta n(\omega)/n_r$ changes sign at the resonance and shifts together with the absorption peak, phase-sensitive or pump-probe measurements can be used to monitor the same geometric control mechanism from a different optical observable \cite{ChemlaShah2001Nature,Dinu2003APL,Autere2018AdvMater}. In this way, absorption and refraction provide two mutually reinforcing probes of the same underlying physics: the first tracks the dissipative part of the response, while the second resolves its dispersive counterpart. The photoionization cross section adds an additional spectroscopic channel, especially useful because it preserves the same state-selective asymmetry while responding differently to changes in photon energy and dipole overlap.

Overall, the present results identify a set of experimentally relevant fingerprints through which torsion, screw dislocation, and longitudinal quantization may be inferred from optical measurements. A systematic observation of (i) blueshift or redshift of resonances, (ii) suppression or enhancement of peak amplitudes, (iii) strong asymmetry between angular-momentum-resolved channels, and (iv) nonlinear bleaching-to-gain evolution under intense excitation would constitute direct evidence of geometry-controlled light--matter interaction in this mesoscopic platform. This makes the system studied here especially attractive as a spectroscopic testbed for identifying and quantifying defect- and torsion-induced confinement in semiconductor nanostructures and related architected media.

\section{Conclusions and Outlook}
\label{conclusion}

In this work, we developed an exact analytical description of a mesoscopic electron system embedded in a helically twisted geometric background and subjected simultaneously to an axial magnetic field and an Aharonov--Bohm flux. Our results show that the combined action of uniform torsion and longitudinal motion generates an effective in-plane harmonic confinement, so that bound states arise from geometry itself, without the need for an externally imposed radial trapping potential. In this sense, torsion serves as a genuine confinement actuator, while the screw dislocation acts as a topological control parameter via a shift in the effective angular momentum.

This geometric mechanism has direct and measurable consequences for the optical response. We demonstrated that increasing torsion enhances effective confinement, blueshifts interlevel transitions, and compresses radial wave functions, thereby reducing dipole overlap and suppressing the amplitudes of the optical resonances. In contrast, the screw dislocation does not create confinement, but redistributes the centrifugal barrier and breaks the dynamical symmetry between positive- and negative-angular-momentum channels. As a result, transitions obeying the same dipole selection rule may occur at very different photon energies and with markedly different oscillator strengths, leading to a strongly asymmetric and state-resolved absorption spectrum.

Within the nonlinear regime, the third-order contribution becomes a decisive ingredient. For sufficiently intense driving, it overcomes linear absorption near resonance and drives the system into a regime of negative absorption, signaling a gain-like optical response. This effect is not universal across all channels: rather, it is selectively controlled by the geometric background. Torsion predominantly governs the spectral position and overall strength of the resonances through confinement-induced squeezing, whereas the screw dislocation can act either as a spectral tuner or an amplitude modulator, depending on the angular-momentum channel involved. The quantum-wire length provides an additional control knob by changing the longitudinal momentum and, therefore, the strength of the torsion-induced confinement. Longer wires weaken the geometric trapping, redshift the resonances, and can strongly enhance the nonlinear gain response.

The photoionization cross section and oscillator strength reinforce this interpretation. Both quantities preserve the same state-selective asymmetry found in the bound-bound optical spectra and provide complementary signatures of the competition between geometric squeezing and dipole overlap. Altogether, our results establish a coherent physical picture in which torsion, topological defect, and longitudinal quantization work together to define the spectral location, amplitude, and nonlinear character of the optical transitions.

These findings indicate that defect and torsion engineering can serve as practical tools for designing geometry-controlled nanophotonic elements, including tunable absorbers, optical switches, and selective gain channels in the mid-infrared and terahertz domains. As natural extensions of this work, it would be interesting to investigate multilevel and many-body effects, excitonic corrections, spin-dependent interactions, and time-dependent driving protocols, and to explore experimental realizations in semiconductor nanowires, strained heterostructures, and architected metamaterials, where torsional and dislocation backgrounds can be effectively tailored.

\section*{Acknowledgments}
This work was partially supported by the Brazilian agencies CAPES, CNPq, FAPEMA and FAPESB. EOS acknowledges the support from grants CNPq/306308/2022-3,
FAPEMA/UNIVERSAL-06395/22, and CAPES/Code 001.

\end{document}